\documentclass[aps,pre,showpacs,12pt]{revtex4-1}
\usepackage{graphicx}
\usepackage{xcolor}
\usepackage{bm,amsmath,amssymb}

\usepackage{subcaption}
\usepackage[percent]{overpic}

\usepackage[T1]{fontenc}
\usepackage[utf8]{inputenc}   
\usepackage{caption}
\usepackage{ragged2e}
\begin{document}
\title{On the thermal properties of knotted block copolymer rings }
\author{Neda Abbasi Taklimi$^1$}
\email{neda.abbasi\_taklimi@phd.usz.edu.pl}
\author{Franco Ferrari$^1$}
\email{franco@feynman.fiz.univ.szczecin.pl}
\author{Marcin Rados{\l}aw Pi\c{a}tek$^1$}
\email{marcin.piatek@usz.edu.pl}
\author{Luca Tubiana$^{2.3}$}
\email{luca.tubiana@unitn.it}
\affiliation{$^1$CASA* and Institute of Physics, University of Szczecin,
  Szczecin, Poland} 
\affiliation{$^2$ Physics Department, University of Trento, Via
  Sommarive 14, I-38123, Trento, Italy}
\affiliation{$^3$ INFN-TIFPA, Trento Institute for Fundamental Physics and Applications, I-38123 Trento, Italy}
\date{\today}

\begin{abstract}
We investigate the thermal and structural properties of knotted diblock
copolymer rings using a coarse-grained lattice model in an implicit
solvent. The system is studied by means of the Wang--Landau Monte Carlo
algorithm, allowing us to analyze thermodynamic and conformational
responses over a wide temperature range. Different knot topologies,
including the unknot, trefoil, figure-eight, and pentafoil knots, are
considered for both symmetric and asymmetric monomer compositions.

In the AB model employed here, A-type monomers are self-repulsive, B-type monomers are self-attractive, and A-B interactions are neutral, such that the solvent is effectively good for A-type monomers and poor for B-type monomers at low temperatures. We analyze several key observables, including the heat capacity, the radius of gyration, and its temperature derivative for both the entire copolymer ring and the individual blocks, and the probability that a monomer belongs to the knotted region. Our results show that the interplay between knot topology, monomer composition, and temperature strongly influences polymer conformations.
Small variations in the B-block length induce nonmonotonic, reentrant-like conformational behavior as a function of temperature, including transitions between knot localization and delocalization at low temperatures.
These effects arise from the competition between energetic and entropic contributions imposed by topological constraints.
\end{abstract}
\maketitle
\section{Introduction}\label{introd}
Knotted polymers have attracted considerable interest from different perspectives~\cite{Tubiana2024, Fielden, Horner, Lim,  Micheletti, Meluzzi, Smerk, Everaers2004, Kauffman2001, Chen2016, Janse2011}.
Controlling and manipulating knots and links within macromolecules at the nanoscale level
provides exciting opportunities for designing novel materials with tailored properties~\cite{Sauvage, Ashbridge, Laurent, Ayme, Lukin, Fan}.
Natural polymers such as DNA and proteins can form knotted structures in living organisms,
which influence their properties and function~\cite{Wasserman, Perlińska, Forte, Taylor, Faísca, Dabrowski, marenduzzo, mallam2006, Jackson2020, Majumder2021}.\\
The effects of nontrivial topology in linear and ring homopolymers have been extensively studied~\cite{Tubiana, Orlandini, Zifferer, Soh, Marcone, Deguchi, Caraglio, Kamenetskii, Grzyb2025, Narros}, as well as in copolymer melts~\cite{herschberg2021} and in the self-assembly of diblock
copolymers into micelles~\cite{Grest, Wil}. 
The coil–globule transitions of cyclic homopolymers and block copolymers with different topologies have been studied under a variety of interaction models and solvent conditions~\cite{ Zhao,Virnau, Wang, Kuriata, Taklim, Taklimi}. 
In a previous study, we investigated the thermal and structural transitions of a knotted figure-eight copolymer ring and a circular poly[4]catenane composed of four diblock copolymers using the contact map matrix method. We demonstrated that the results obtained from this method are consistent with the corresponding specific heat capacity signatures, including peaks associated with structural transitions~\cite{Taklimi}.  

The main motivation of this work is to understand how variations in monomer composition, topology (knot type), and the strength of monomer interactions influence the thermal and structural behavior of a single diblock copolymer ring. 
Tagliabue \emph{et al.}~\cite{Tagliabue21,Tagliabue22} reported that in knotted copolyelectrolyte rings composed of charged and neutral segments, knot size and localization can be tuned by varying the length of the neutral block. They showed that electrostatic interactions, mediated by counterions, salt concentration, and solvent properties, play a key role in this behavior. They found that the knot size and conformational properties exhibit a nonmonotonic dependence on the neutral block length, accompanied by a transition in the knot position from pinning at the edge of the neutral segment to localization within it. Similar effects have been reported for polyampholytes, where charge distribution influences knot dynamics and stability; in particular, the knot lifetime increases with block length in alternating charge sequence~\cite{Ozmaian}. Moreover, stiffness heterogeneity along the chain can control both the knot length and its position in diblock ring polymers~\cite{Enzo,copoknotP}. 

Overall, previous studies have established that charge heterogeneity and chain stiffness govern knot size, position, and conformational properties in diblock ring copolymers. In the present work, we demonstrate how knot type and localization influence heat capacity and temperature-dependent conformational transitions in an implicit solvent.

 We consider a single knotted diblock copolymer ring on a cubic lattice, where the relative lengths of the two blocks are systematically varied by slightly enlarging one block while shortening the other. 
To investigate this system, simulations are performed using the Wang--Landau Monte Carlo algorithm~\cite{Wang2001}, which estimates the density of states via a random walk in energy space and enables the calculation of the expectation values of the observables in the canonical ensemble at any temperature range. We first determine the energy range accessible to the system and then perform simulations within this interval. A precise determination of this range is essential in order to consider the structural transition at low temperatures.
To explore the influence of interaction strength on the physical properties, we employ the AB model, in which interactions between A-type monomers are repulsive, while those between B-type monomers are attractive. Interactions between unlike monomers consist only of excluded-volume effects.
The solvent can therefore be regarded as a good solvent for A-type monomers and a poor solvent for B-type monomers, leading to swelling and collapse at low temperatures, respectively. Equivalently, the system can be viewed as a copolymer ring in solution with hydrophilic and hydrophobic interactions.
The HP polymer folding model, in which the interactions of the A block are switched off, is also used as a benchmark for comparison.

To characterize the structural arrangements of the knotted copolymer ring in a given phase, we analyze the sampled polymer conformations using the package KymoKnot~\cite{tubiana2018kymoknot}, which provides the minimal length of the knotted segment and its position along the ring. 
The results are then averaged using the density of states obtained via the Wang–Landau method, enabling the estimation of the probability that a given bead belongs to the knotted region of the ring at a specified temperature. Owing to the computational cost of knot detection, the analysis is limited to a subset of configurations at each energy, and the resulting quantities should be regarded as qualitative estimates.

Remarkably, we find that in the AB model, introducing only a few attractive B-type monomers ($N_A > N_B$) can substantially alter the system’s structural transitions. This sensitivity reflects the competition between energetic and entropic contributions to the free energy, which governs the localization or delocalization of the knot and thereby determines the system’s stability. Within this framework, distinct regimes emerge in which knot localization is associated with nonmonotonic conformational behavior.

The remainder of this paper is organized as follows. In Sec.~\ref{method}, we describe the methodology. The thermal properties of knotted copolymer rings for short and long chain lengths are presented in Secs.~\ref{r}A and~\ref{r}B, respectively. A comparison between the HP and AB models, as well as the effect of knot topology on polymer size, is discussed in Secs.~\ref{r}C and~\ref{r}D. Finally, conclusions are presented in Sec.~\ref{con}.

\section{Methodology}\label{method}
Coarse-grained models are widely used to capture the essential physical properties of polymer systems while
significantly reducing the computational cost compared to more detailed simulations~\cite{Müller}. We consider a ring polymer in an
implicit solvent, where effective monomer–monomer interactions account for its behavior
in a solution. Initial knotted and unknotted configurations are constructed as self-avoiding rings on a three-dimensional simple cubic lattice. These configurations are generated from minimal-length knotted conformations of closed loops with different topologies~\cite{van,Scharein} and subsequently extended to the desired chain lengths \(N=90\) and \(200\).

\begin{figure}[t]
\centering
\includegraphics[width=0.45\textwidth]{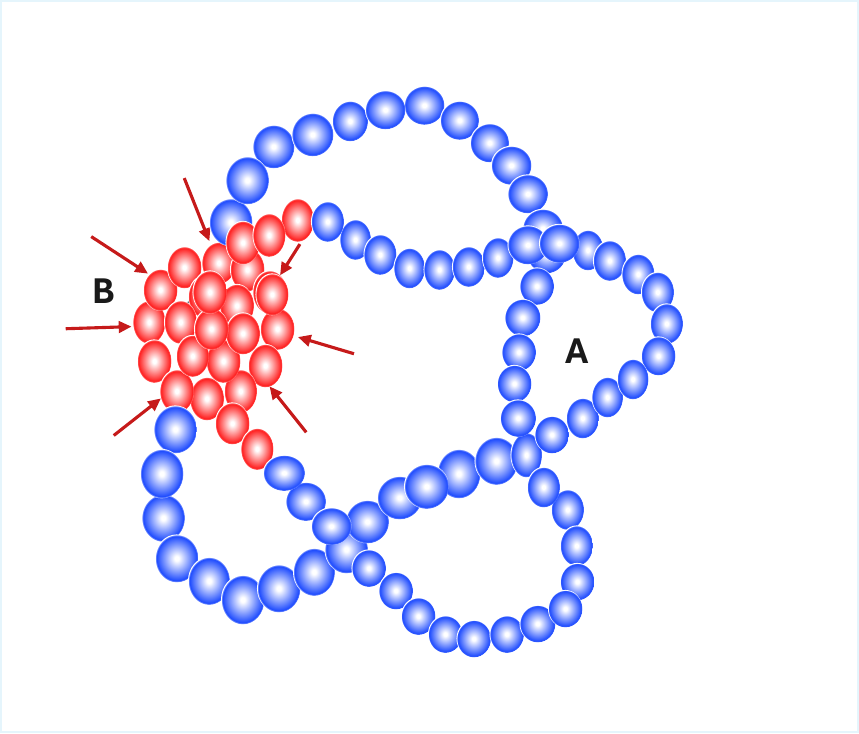}
\caption{\justifying Schematic representation of the coarse-grained AB model of a knotted diblock copolymer ring in a solution. A-type monomers (blue) are solvophilic and mutually repulsive, whereas B-type monomers (red) are solvophobic and attract each other, forming a collapsed domain at low temperatures. Interactions between A and B monomers are neutral.}
\label{model}
\end{figure}

The total number of monomers in the copolymer ring is \(N = N_A + N_B\), where \(N_A\) and \(N_B\) denote the numbers of A- and B-type monomers, respectively. Short-range interactions proportional to the number of contacts between monomers of each type are considered. The Hamiltonian \(H(X)\) of a given polymer conformation \(X\) is defined as

\begin{equation}
H(X) = \varepsilon \left( m_{AA} - m_{BB} \right)
\label{hamII}
\end{equation}

where \( m_{MM'} \) denotes the number of non-bonded contacts between monomers of types \( M \) and \( M' \), with \( M, M' = A, B \). Two monomers \( i \) and \( j \) are considered to be in contact if \( i \neq j \pm 1 \) and \( |\boldsymbol{R}_i - \boldsymbol{R}_j| = 1 \), where \( \boldsymbol{R}_1, \ldots, \boldsymbol{R}_N \) denote the lattice positions of the \( N \) monomers. The signs of the interaction terms in Eq.~\eqref{hamII} imply that A--A interactions are repulsive, whereas B--B interactions are attractive. Interactions between unlike monomers (A--B) are neutral and contribute only through excluded-volume constraints. Consequently, A-type monomers tend to swell and avoid mutual contacts, while B-type monomers favor aggregation to minimize the free energy of the system.
Introducing the interaction parameters \( \varepsilon_{A\text{-}A} = 1 \) and \( \varepsilon_{B\text{-}B} = -1 \), the Hamiltonian can be written in reduced form as
\begin{equation}
H^*(X) = \varepsilon_{A\text{-}A} m_{AA} + \varepsilon_{B\text{-}B} m_{BB} \, .
\label{hamII2}
\end{equation}
We define the reduced temperature as
\begin{equation}
T^* =\frac{k_BT}{\varepsilon},
\end{equation}
with the Boltzmann constant $k_B$ set to unity (thermodynamic units).
This Hamiltonian describes the behavior of a single diblock copolymer in a selective solvent that is good for A-type monomers and poor for B-type monomers. A schematic representation of the model is shown in Fig.~\ref{model}. Energetic interactions and topology-dependent entropic constraints jointly determine the equilibrium conformations.
A major challenge in sampling the conformations of knotted polymers is preserving their topology during simulations while maintaining computational efficiency. Although several topological invariants, such as the Alexander and Jones polynomials~\cite{Micheletti}, can be used to distinguish different knot types, their evaluation is computationally too expensive to be performed during sampling. To efficiently preserve and monitor the topology of knotted polymers, a topology-preserving Monte Carlo approach based on the pivot algorithm with excluded volume area (PAEA) is employed~\cite{Zhao}. Within this framework, new configurations are generated while preserving the topological state of the polymer chain.

The pivot algorithm employed in this work consists of three types of moves: interchange, reflection, and inversion~\cite{Madras, Lai}. A proposed move is accepted only if it preserves self-avoidance and does not introduce intersections within the loop formed between the initial and transformed segments. Each Monte Carlo move acts on a short segment of the polymer chain, randomly selected to contain between three and five monomers. Owing to its nonlocal updates, the pivot algorithm efficiently samples the configurational space, but it does not represent physical polymer dynamics~\cite{Madras}. Further details regarding the sampling procedure and the implementation of topological constraints within the PAEA method can be found in Ref.~\cite{Ferrari2017}.

To efficiently sample the free energy landscape, we adopted the Wang-Landau method, a flat-histogram approach that performs a random walk in energy space to estimate the density of states \( g(E) \). A key advantage of this method is that it provides direct access to \( g(E) \), defined as the number of configurations corresponding to a given energy, thereby enabling the calculation of canonical-ensemble expectation values of arbitrary observables over a wide range of temperatures.

The partition function of the polymer ring is given by
\begin{equation}
Z(T^*) = \sum_{E=E_{\min}}^{E_{\max}} g(E)\, e^{-E/T^*},
\end{equation}
where \( g(E) \) is the density of states, defined as
\begin{equation}
g(E) = \sum_X \delta\!\left(H(X)-E\right).
\end{equation}
Here, \(E_{\min}\) and \(E_{\max}\) denote the minimum and maximum energies, respectively, determined from a preliminary simulation run performed without imposing any energy bounds. This preliminary run is terminated once no new energy values are sampled. Subsequently, a second simulation is carried out to compute microcanonical ensemble averages using the values of \(E_{\min}\) and \(E_{\max}\) obtained from the preliminary run. In this second run, the energy range remains unrestricted to check if new energy values are discovered. If this is the case, the run is restarted with the new energy boundary. The convergence of the Wang--Landau algorithm is assessed only within the interval \([E_{\min}, E_{\max}]\). This requires the sampling of a number of the conformations of the order of \(10^{12}\).

The observables considered in this work include the mean specific heat capacity \(C_V(T^*)\), the radius of gyration \(R_g^2\), and the contributions from the individual blocks of the ring polymer. The expectation value of an observable \(\mathcal{O}\) is computed using the formula:
\begin{equation}
\langle \mathcal{O} \rangle(T^*) =
\frac{1}{Z(T^*)}
\sum_{E=E_{\min}}^{E_{\max}}
g(E)\, e^{-E/T^*}\, \mathcal{O}_E,
\label{4}
\end{equation}
where \(\mathcal{O}_E\) denotes the average of \(\mathcal{O}\) over all sampled configurations with energy \(E\).
The canonical mean energy and the specific heat capacity per monomer are given by Eqs. (5) and (6), respectively.
\begin{equation}
\langle E(T^*) \rangle
=
\frac{1}{Z(T^*)}
\sum_{E=E_{\min}}^{E_{\max}} E\, g(E)\, e^{-E/T^*},
\label{en}
\end{equation}

\begin{equation}
\frac{C_V(T^*)}{N}
=
\frac{\langle E^2(T^*) \rangle - \langle E(T^*) \rangle^2}
{N {T^*}^2},
\end{equation}
Since there are no energetic interactions between A- and B-type monomers and the Hamiltonian consists of pairwise contact interactions [see Eq.~\eqref{hamII}], the energetic contributions of the two blocks can be evaluated separately as
\begin{equation}
\langle E(T^*) \rangle
=
\varepsilon \left[
\langle m_{AA}(T^*) \rangle
-
\langle m_{BB}(T^*) \rangle
\right].
\end{equation}
Using the thermodynamic relation
\begin{equation}
C_V(T^*) = \frac{d\langle E(T^*) \rangle}{dT^*},
\end{equation}
the specific heat capacity can be decomposed into contributions from each block,
\begin{equation}
C_V(T^*) = C_{AA}(T^*) + C_{BB}(T^*),
\end{equation}
with
\begin{equation}
C_{AA}(T^*) = \varepsilon \frac{d\langle m_{AA}(T^*) \rangle}{dT^*},
\qquad
C_{BB}(T^*) = -\varepsilon \frac{d\langle m_{BB}(T^*) \rangle}{dT^*}.
\end{equation}

Analogously, the total radius of gyration of a diblock copolymer can be written as the sum of three terms,
\begin{equation}
R_g^2 =
\frac{N_A}{N} R_{g,A}^2
+
\frac{N_B}{N} R_{g,B}^2
+
\frac{N_A N_B}{N^2}
\left|\mathbf{r}_{\mathrm{cm},A} - \mathbf{r}_{\mathrm{cm},B}\right|^2 ,
\label{gyr}
\end{equation}
where the first two terms describe the contributions of blocks~A and~B, respectively, while the third term accounts for the separation between their centers of mass~\cite{Fredrickson, Rubinstein}.

The energetic and structural behavior described above is governed by the interaction scheme of the AB model. The AB Hamiltonian is a variant of the HP (hydrophobic--polar) protein-folding model~\cite{dill}, in which attractive interactions between hydrophobic monomers drive chain compaction. The HP model serves as a reference to assess the effect of repulsive interactions introduced in the AB model on the system’s physical properties. 
In the HP model, only attractive interactions between hydrophobic (B-type) monomers are considered. Accordingly, Eq.~(1) reduces to
\[
H_{\mathrm{HP}}(X) = -\epsilon m_{BB},
\]
where $m_{BB}$ denotes the number of B--B contacts.

Diblock copolymer rings in a given monomer composition, consisting of  $N_A$ monomers of type A and $N_B$ monomers of type B, are denoted by $\mathrm{K}(N_A, N_B)$, where
$\mathrm{K}$ specifies the topological state of the ring, chosen from the
unknot $0_1$, trefoil $3_1$, figure-eight knot $4_1$, and pentafoil knot
$5_1$. Knotted homopolymer rings in good and
poor solvents are denoted by $\mathrm{K}(\mathrm{H}(+))$ and
$\mathrm{K}(\mathrm{H}(-))$, respectively. We further define
$f = N_B/N$ as the fraction of B monomers in the copolymer ring. Symmetric and asymmetric copolymer compositions are characterized by $N_A \approx N_B$ and $N_A \neq N_B$, respectively.
\section{RESULTS } \label{r}
\subsection{THERMAL PROPERTIES OF SHORT KNOTTED DIBLOCK COPOLYMER RINGS ( N=90 )}~\label{short}
\begin{figure}
  \begin{center}
    \includegraphics[width=0.48\textwidth]{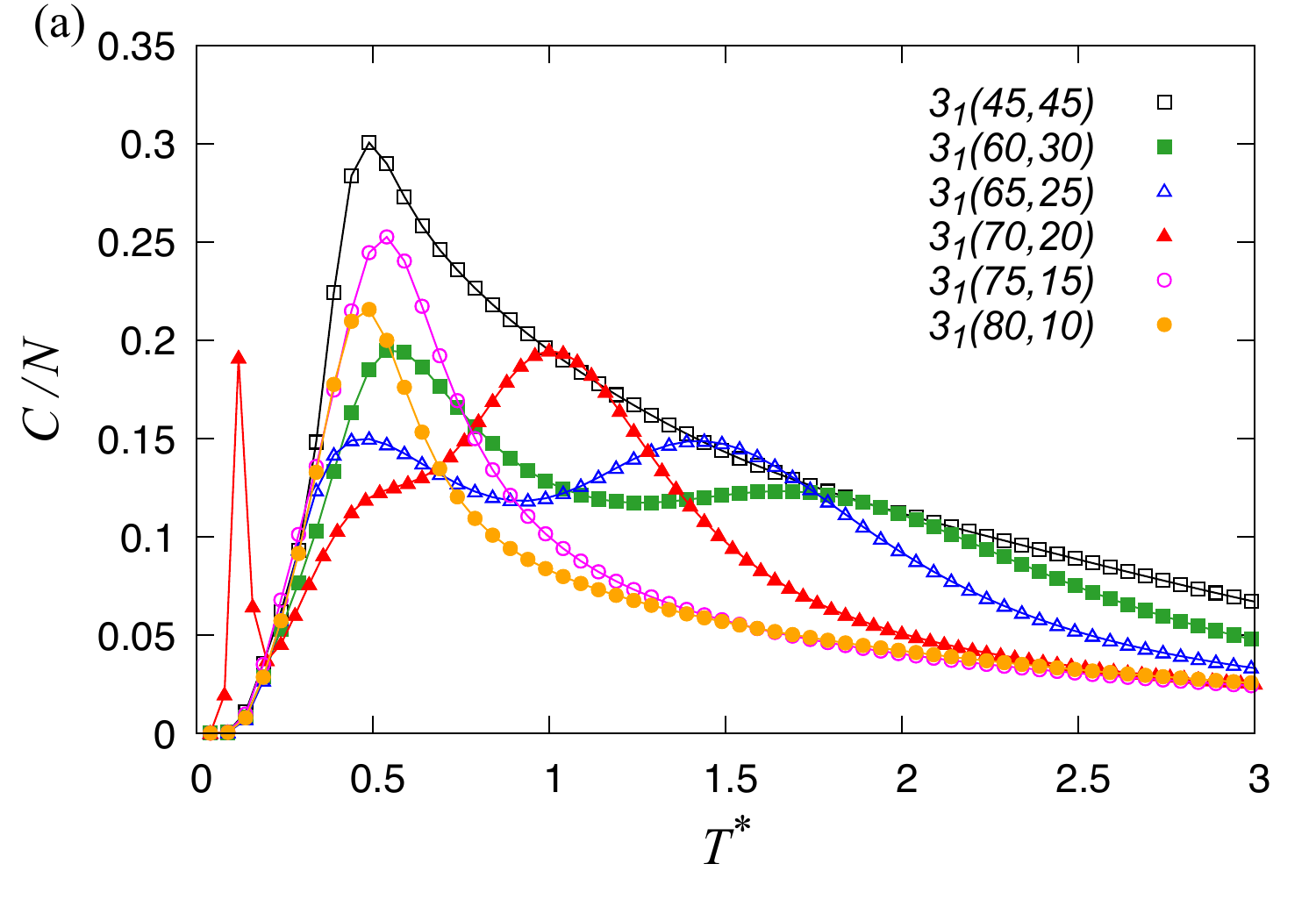}
     \includegraphics[width=0.48\textwidth]{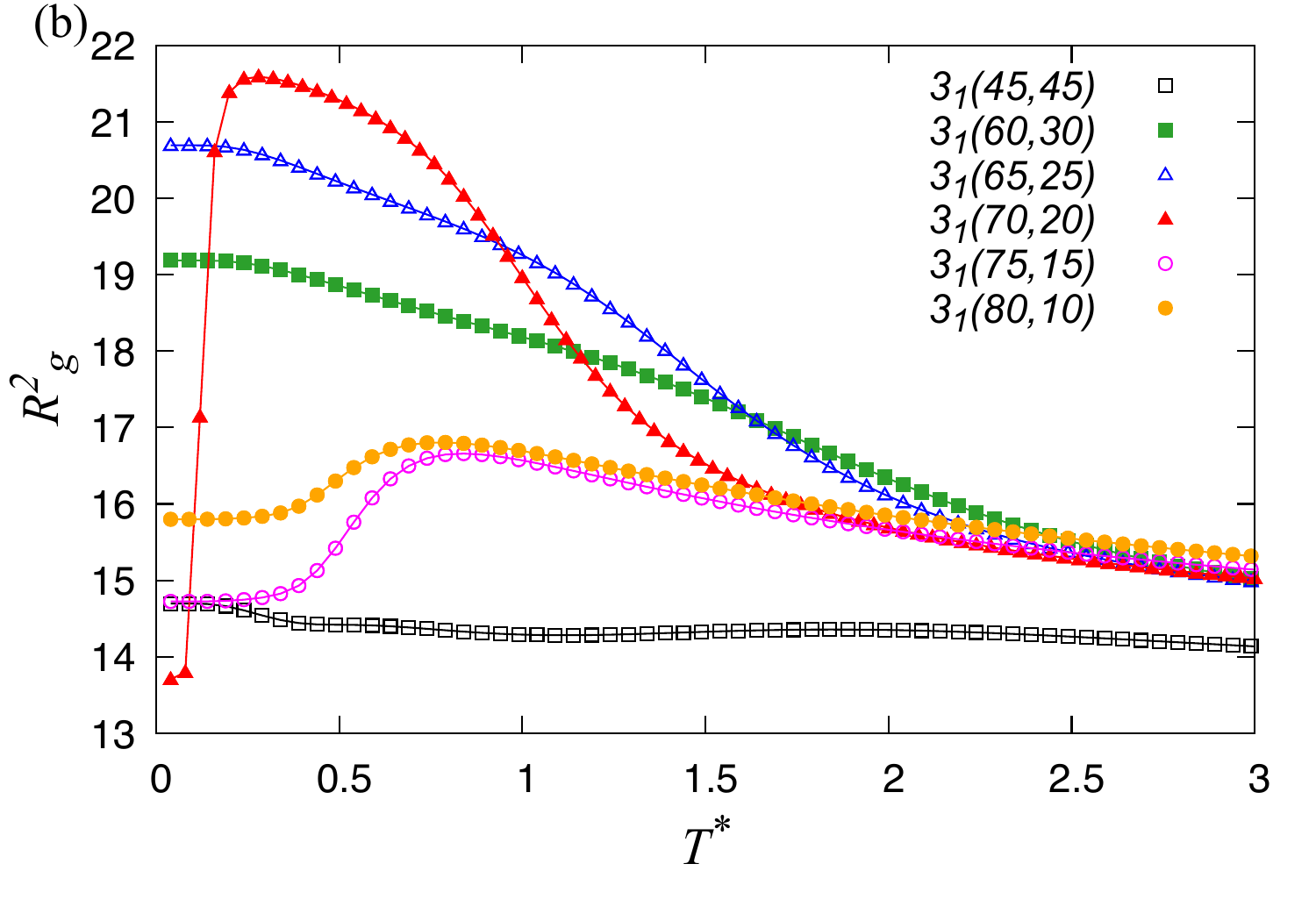}
         \includegraphics[width=0.48\textwidth]{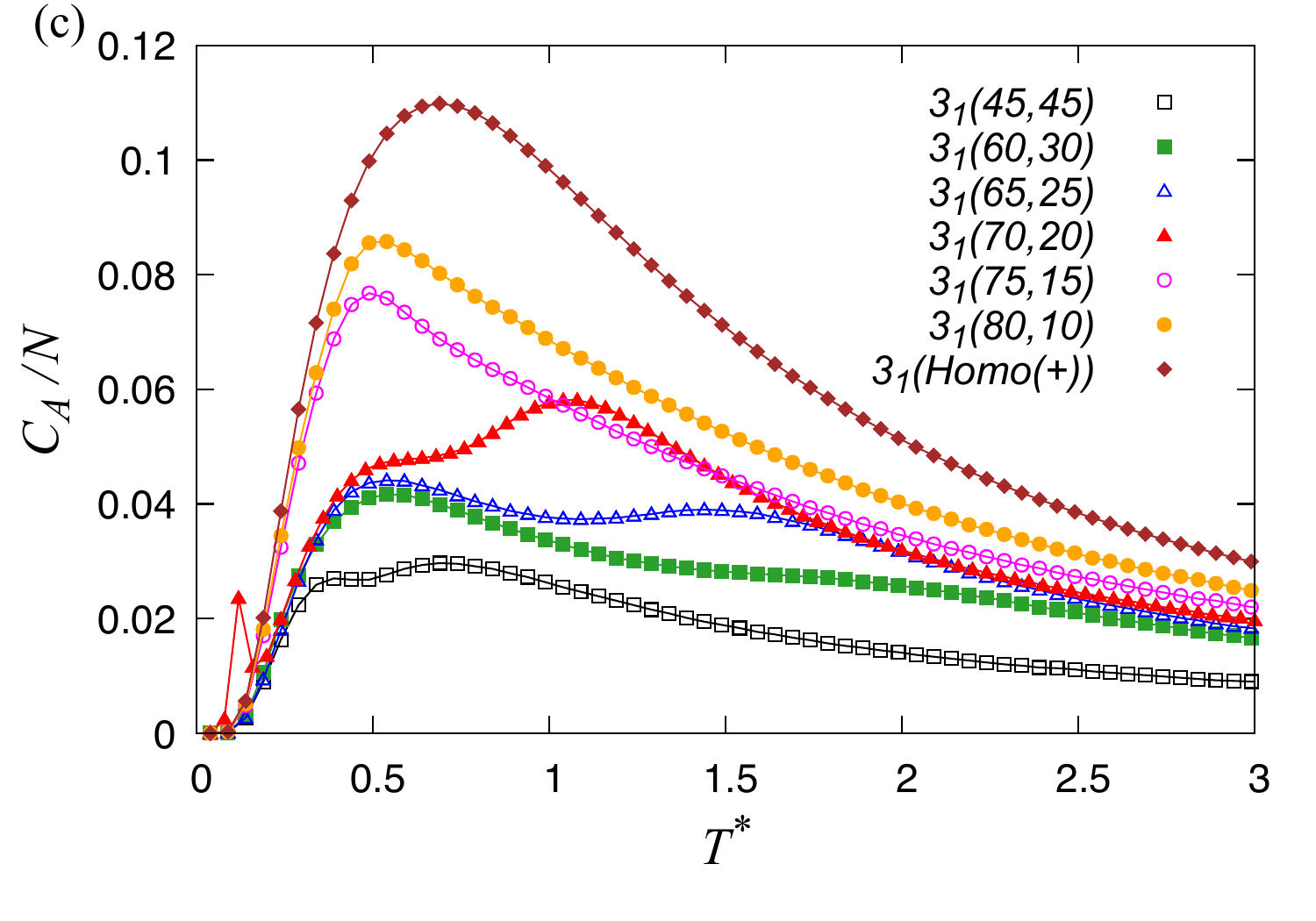}
     \includegraphics[width=0.48\textwidth]{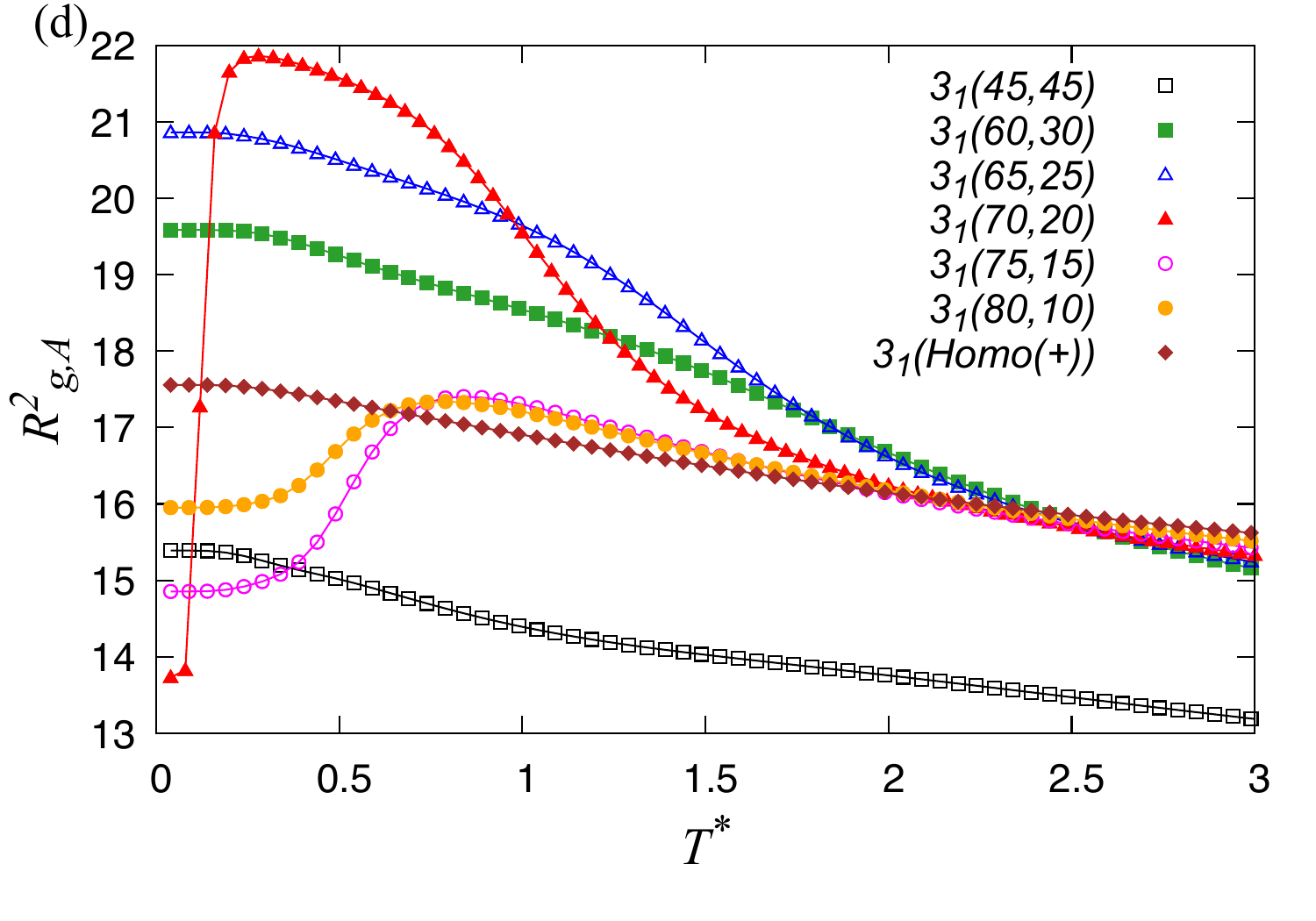}
      \includegraphics[width=0.48\textwidth]{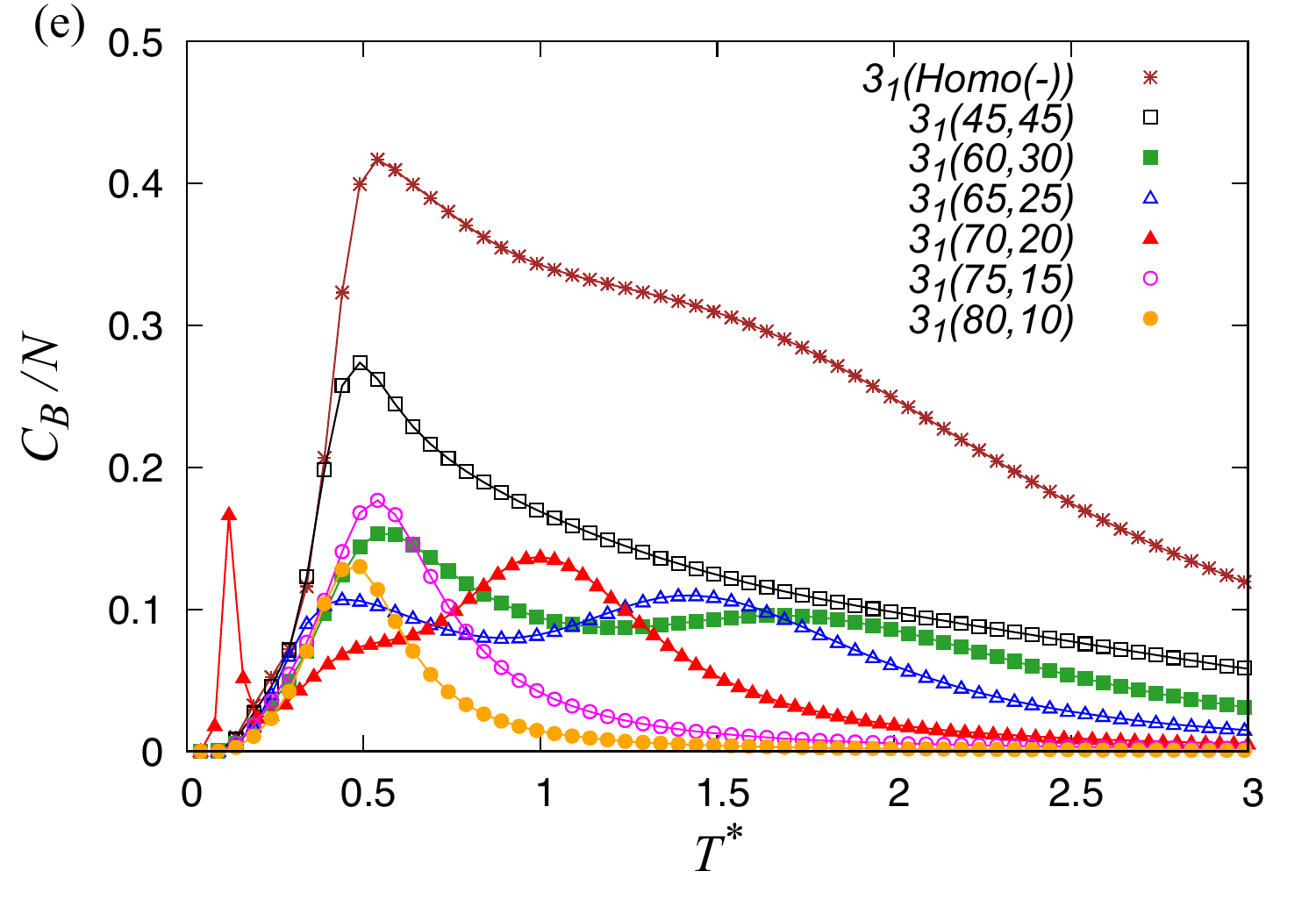}
    \includegraphics[width=0.48\textwidth]{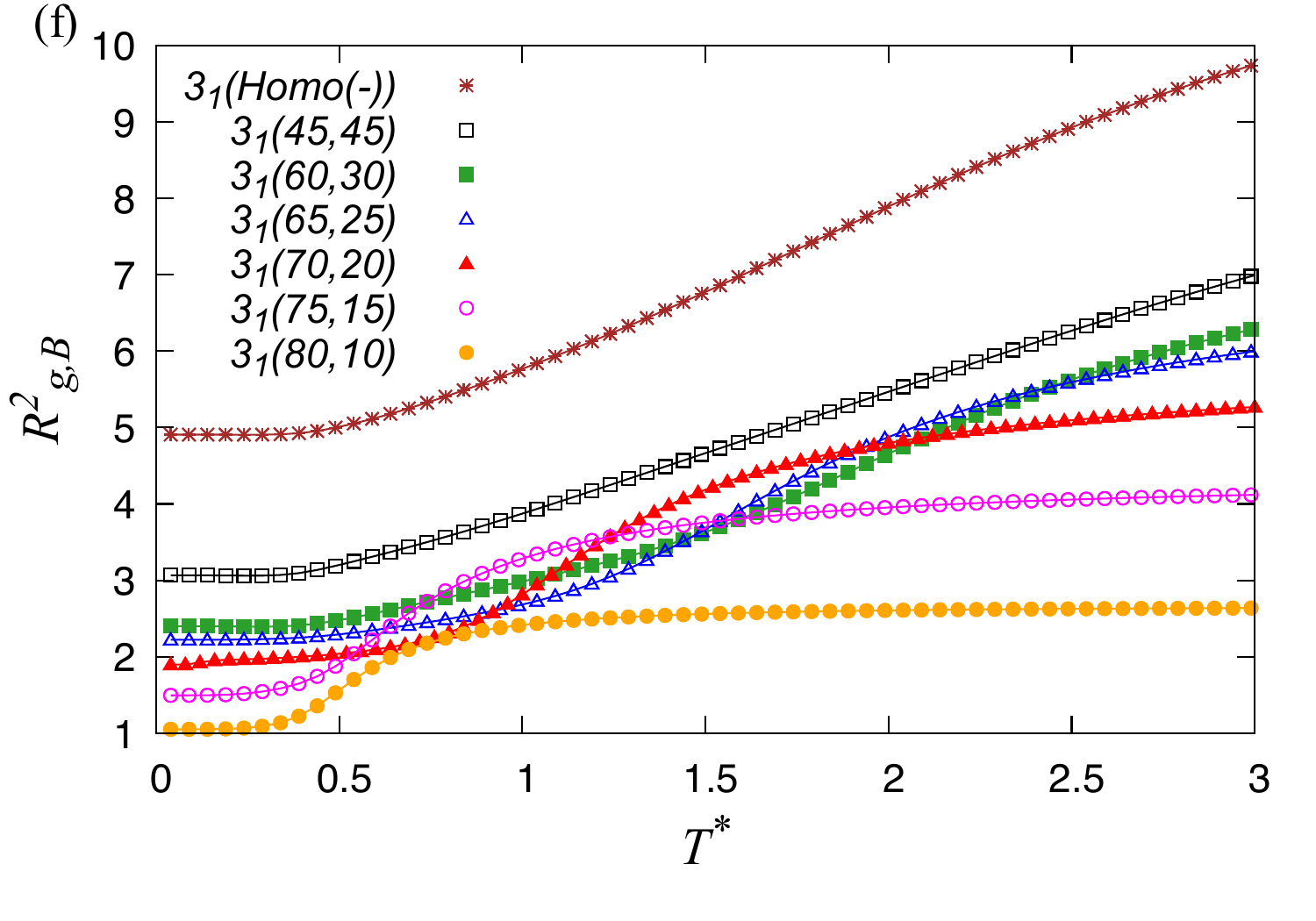}     
    \caption{\justifying Thermodynamic and structural properties of knotted diblock copolymer rings with topology $3_1$ and total length $N=90$ for different monomer compositions. 
The panels represent: (a) specific heat capacity $C/N$ as a function of reduced temperature $T^*$,
(b) mean-square radii of gyration $R_g^2$,
(c) specific heat capacities of the A block, normalized by the total number of monomers $N$,
(d) mean-square radii of gyration of the A block,
(e) specific heat capacities of the B block, normalized by the total number of monomers $N$, and
(f) mean-square radii of gyration of the B block.}
      \label{3.1-cv-Rg}       
\end{center}
\end{figure}

We now investigate diblock copolymer rings with symmetric ($N_A \approx N_B$) and asymmetric ($N_A > N_B$) monomer compositions. For short chains with total length $N = 90$, the trefoil ($3_1$) and pentafoil ($5_1$) knots are considered as representative topologies. Fig.~\ref{3.1-cv-Rg} shows the temperature dependence of the heat capacity (panels a, c, and e) and the mean-square radius of gyration (panels b, d, and f) for trefoil polymer rings, for both the entire copolymer ring and the individual blocks.

For asymmetric monomer compositions with $N_B \leq 20$, a nonmonotonic temperature dependence of the mean-square radius of gyration is observed, particularly for $3_1(70,20)$. In this case, at very low temperatures, the polymer adopts a compact conformation, followed by an expansion upon heating and subsequent shrinkage at higher temperatures (see Fig.~\ref{3.1-cv-Rg}(b)).

The corresponding signature of this behavior is also reflected in the heat capacity. The first peak in the heat capacity curve of $3_1(70,20)$, located around $T^* \simeq 0.12$, correlates with the expansion observed in the radius of gyration. This behavior can be attributed to the onset of knot localization within the B block, corresponding to a transition from weak localization at low temperatures to strong localization at intermediate temperatures, governed by the balance between entropic and energetic contributions.
 When the knot is weakly localized, the A block, subjected to short-range repulsive interactions,  becomes more compact due to the topological constraints.
A similar, but progressively weaker, trend in the radius of gyration is observed for the compositions $3_1(75,15)$ and $3_1(80,10)$, although the associated heat capacity curves no longer exhibit a distinct low-temperature peak. In these cases, the smaller size of the B block prevents the knot from being fully accommodated within the B region. Upon heating, the knot can shift toward the B segment, leading to a moderate extension of the A block and a correspondingly weak nonmonotonic structural response. As it is possible to see in Fig.~\ref{3.1-cv-Rg}(b), this moderate extension occurs at a temperature that is much higher than the temperatures of similar extension in the case of $3_1(70,20)$. Consequently, knot relocation and B-block melting become coupled, leading to a single broadened peak in the heat capacity, rather than distinct transitions as observed for $3_1(70,20)$.
\begin{figure}
\centering

\begin{subfigure}{0.48\textwidth}
\centering
\includegraphics[width=\linewidth]{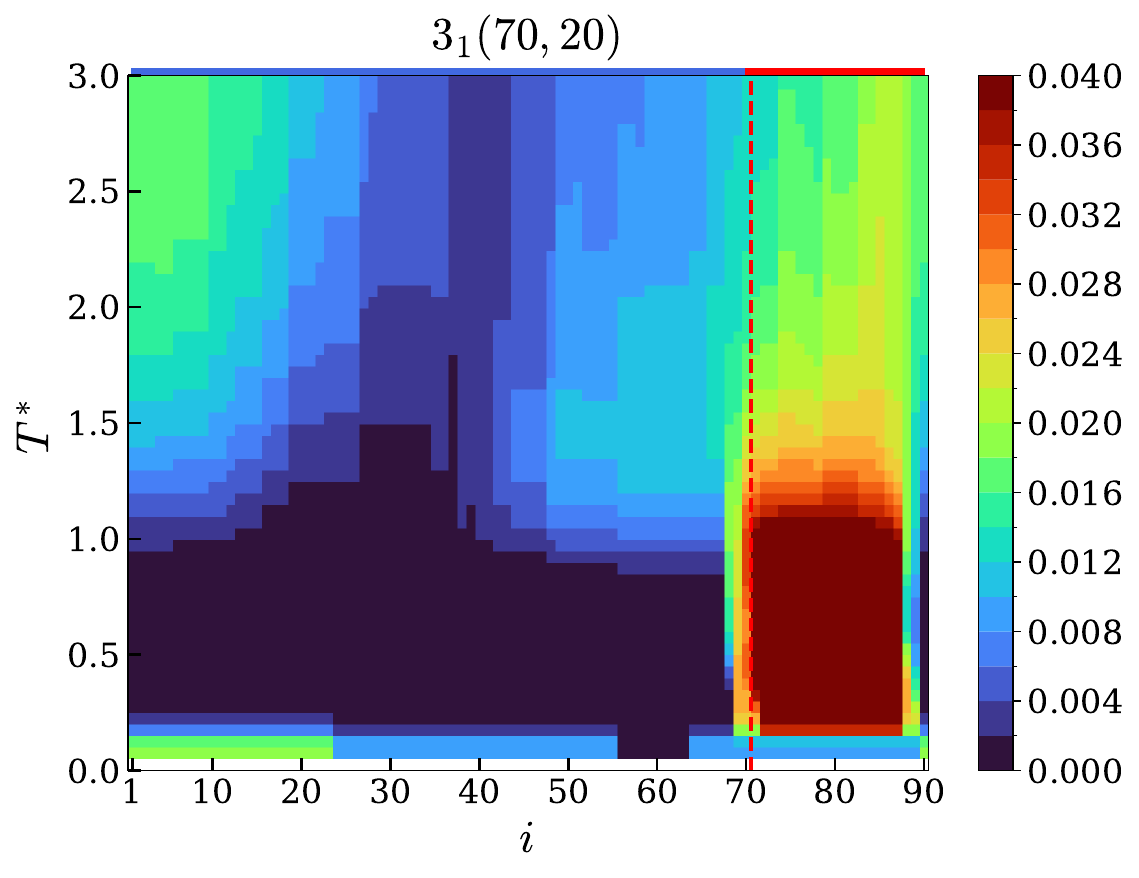}

\vspace{-18pt}
\centering (a)
\end{subfigure}
\hfill
\begin{subfigure}{0.48\textwidth}
\centering
\includegraphics[width=\linewidth]{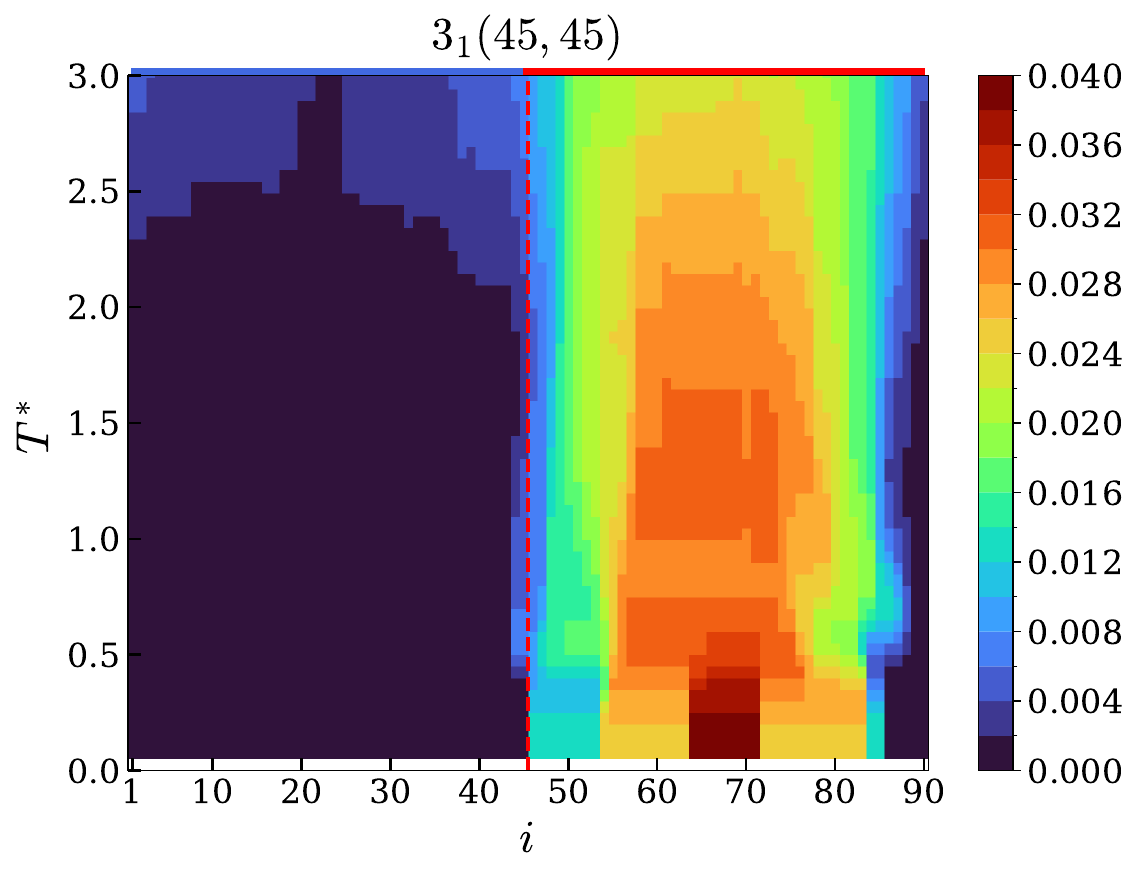}

\vspace{-18pt}
\centering (b)
\end{subfigure}
\caption{\justifying Knot localization maps for the copolymer ring with the trefoil topology ($3_1$) and different monomer compositions: (a) $3_1(70,20)$ and (b) $3_1(45,45)$.The colors indicate the likelihood of monomer $i$ being located within the minimal knotted region as a function of reduced temperature $T^*$ and monomer position along the chain. The vertical line denotes the boundary between the $A$ and $B$ blocks.}
\label{3.1-Pit-N90}
\end{figure}

\begin{figure}[t]
\centering

\begin{subfigure}{0.32\textwidth}
\centering
\begin{overpic}[width=\linewidth]{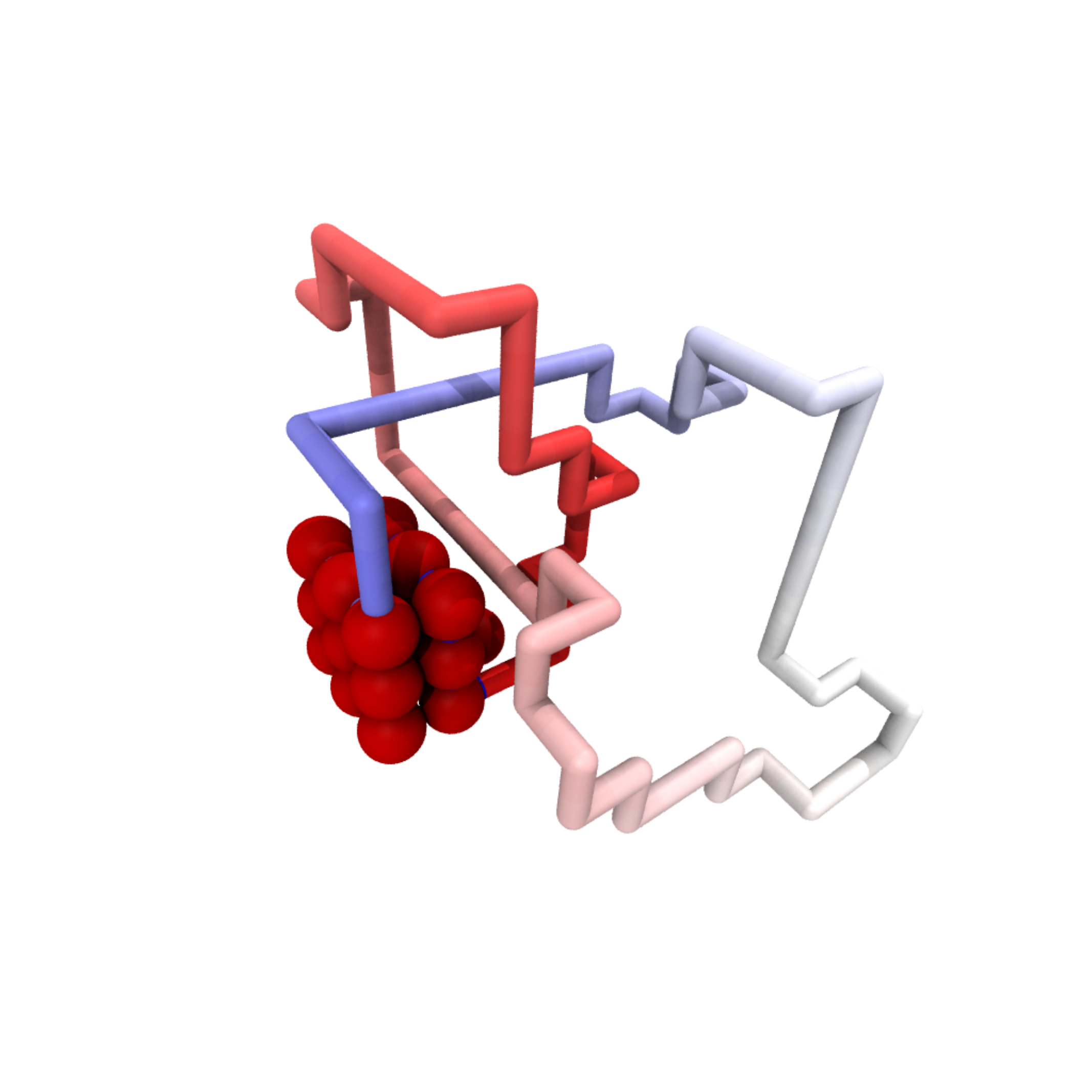}
\put(44,6){\normalsize (a)}
\end{overpic}
\end{subfigure}
\hfill
\begin{subfigure}{0.32\textwidth}
\centering
\begin{overpic}[width=\linewidth]{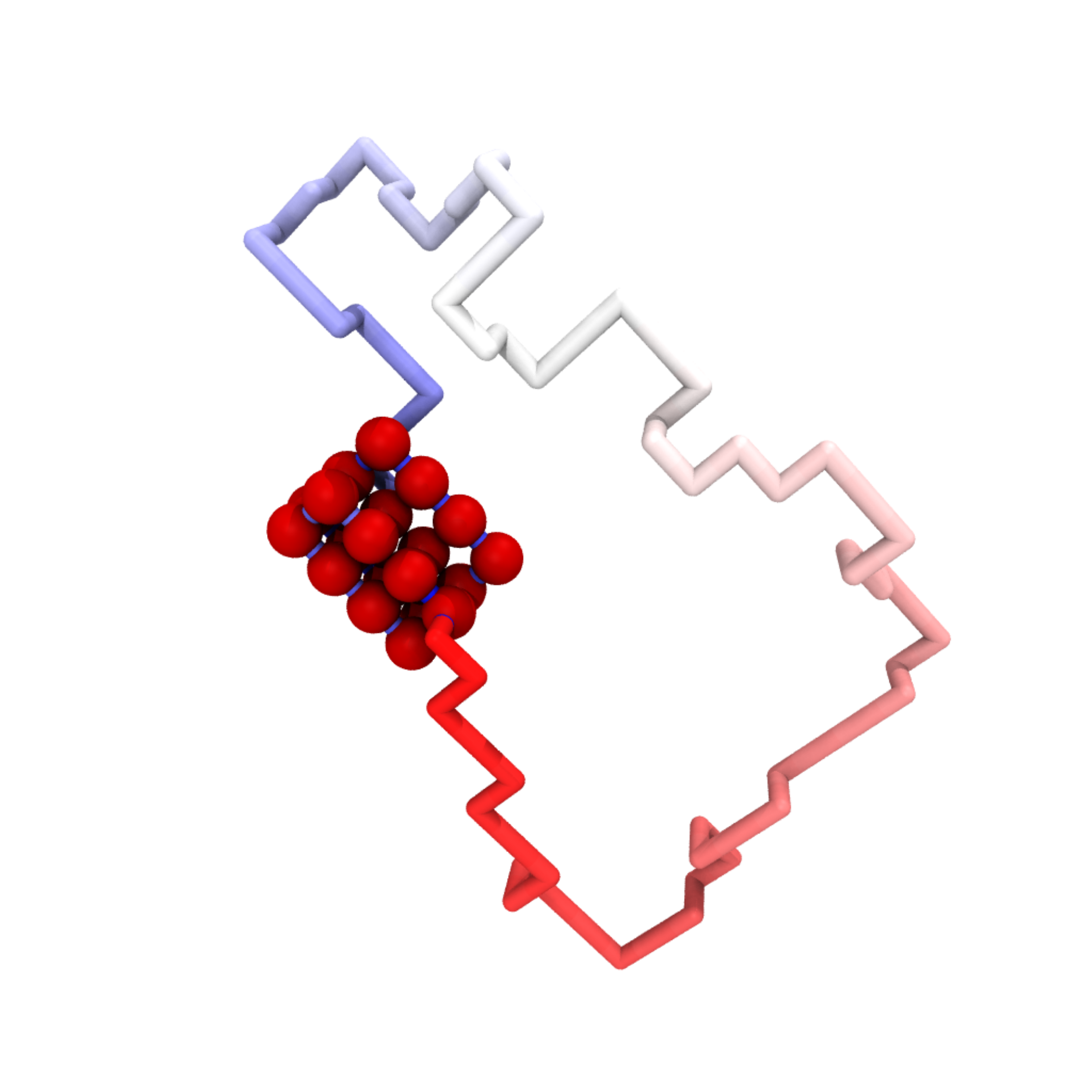}
\put(44,6){\normalsize(b)}
\end{overpic}
\end{subfigure}
\hfill
\begin{subfigure}{0.32\textwidth}
\centering
\begin{overpic}[width=\linewidth]{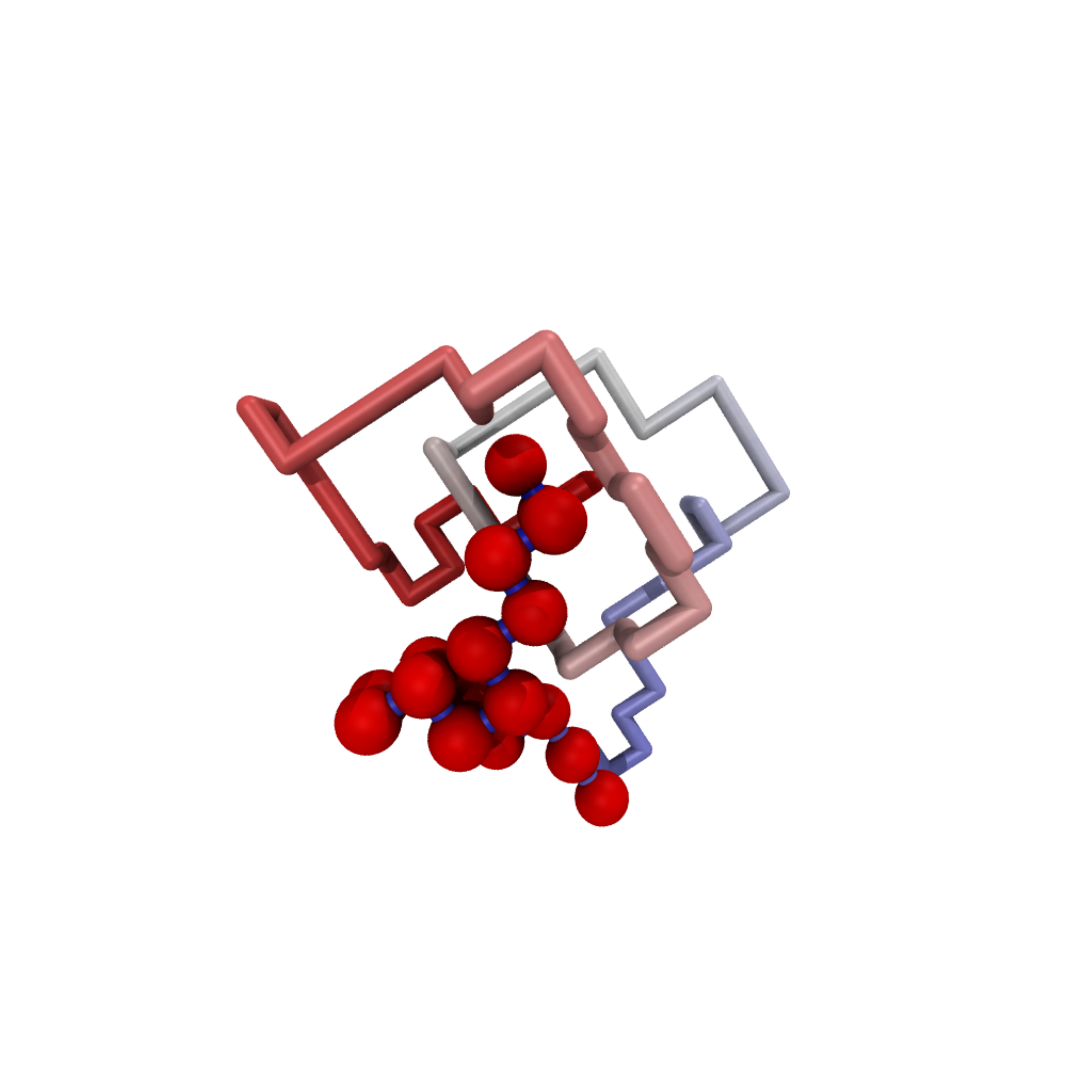}
\put(44,6){\normalsize(c)}
\end{overpic}
\end{subfigure}
\caption{\justifying Typical configurations of the copolymer ring with trefoil topology with monomer composition (70, 20)
(a) the ground state at the lowest sampled energy ($E = -17$),
(b) a configuration at energy $E = -16$, and
(c) a configuration at a higher energy $E = 6$.
The B block is shown as red beads, while the A block is represented by lines.}

\label{confo3_1}
\end{figure}

To characterize the behavior described above, we use knot localization maps, see Fig.~\ref{3.1-Pit-N90}, obtained with the help of the KymoKnot program.
The values reported in the maps provide a qualitative illustration of the
relationship between polymer conformation and knot localization upon heating. More precisely, the color map scales present the likelihood that a given monomer belongs to the
knotted part as defined in Eq.~\ref{4}. The analysis reveals that the knot length and position
play key roles in controlling the overall size of the copolymer ring.

In Fig.~\ref{3.1-Pit-N90}(a), for the case $3_1(70, 20)$, the lowest-temperature state exhibits a delocalized knot extending over a large portion of the ring. This configuration represents the ground state with $E = -17$, as shown in Fig.~\ref{confo3_1}(a). It corresponds to the energy-minimizing conformation of the system, in which the knot extends along the ring rather than localizing within the B block. In this regime, the B monomers collapse into a compact structure, making the B block unfavorable for accommodating the knot.
As the temperature increases, the knot becomes strongly localized within the B segment. The knotted region is highlighted in red in the color map, and the corresponding configuration is shown in Fig.~\ref{confo3_1}(b). This conformation has a slightly higher energy $(E = -16)$, in which the knot becomes fully localized within the B block; consequently, the A block relaxes and adopts a more extended conformation, leading to an overall increase in the size of the $3_1(70, 20)$ ring (see Fig.~\ref{confo3_1}(b)). A change in energy by a single unit is sufficient to induce this structural transition, reflecting a delicate balance of free-energy contributions.
Upon further heating, the B block melts, and the knot progressively delocalizes over the entire ring as indicated by the light blue and green regions in the color
map of Fig.~\ref{3.1-Pit-N90}(a), in agreement with the
observed reduction in the total radius of gyration (see Fig.~\ref{3.1-cv-Rg}(b) and Fig.~\ref{confo3_1}(c)). This behavior originates from the fact that when $N_{B} = 20 $,
the B block is too small to accommodate the knot without significant conformational distortion, resulting
in two competing low-energy states that differ in the degree of knot localization: a weakly localized state in which the knot extends over the ring, and a localized state restricted within the B block. The transition between these states gives rise to a sharp
heat capacity peak. In contrast, for $3_1(65,25)$ with $N_B = 25$, the larger B block can accommodate the
knot without requiring a distinct structural rearrangement, resulting in a smoother energy
landscape, such that the structural changes occur gradually and no sharp heat capacity peak
is observed. In this case, the gyration radius exhibits a comparatively large
overall size at very low temperatures, the knot localizes primarily within the B block, effectively releasing the A block, so that it can adopt a highly extended conformation. As the temperature
increases, the B block begins to melt, and the knot becomes less localized, allowing it to
move along the ring. As a result, the A block becomes considerably restricted, leading to a
reduction in its extension. 

The nonmonotonic dependence of the gyration radius observed in this system is reminiscent of reentrant behavior~\cite{Zhang, Yong, Okay}; however, it does not represent a genuine reentrant transition, as the compact
conformations at low and high temperatures are microscopically distinct. Here, “reentrant-like” denotes a nonmonotonic temperature dependence of the gyration radius and knot
localization in a finite system, without implying a true thermodynamic transition.
The origin of this behavior for B blocks containing a number of the order of 20 monomers
can be related to the minimal number of monomers required to form a trefoil knot, which
defines an optimal block size for knot localization at intermediate temperatures. Optimal
refers to the block size that maximizes the growth of the gyration radius. 
 At lower temperatures, there is another optimal size of the B block that maximizes the gyration radius, and it requires a larger value, around 25 monomers, as indicated in Fig.~\ref{3.1-cv-Rg}(b) by the fact that the gyration radius of the blue line has the maximum value near zero temperature.

For the symmetric diblock copolymer $3_1(45, 45)$, the polymer undergoes moderate shrink-
age with increasing temperature, while the heat capacity exhibits a broad and enhanced
peak. Although $N_{A}= N_{B}$, the center of mass of the copolymer is shifted toward block A
due to its more swollen conformation, leading to unequal contributions of the two blocks to
the total radius of gyration. In particular, for $3_1(45, 45)$, the squared radius of gyration of
block~A, $R^2_{g, A}$, decreases by approximately 2.5 units between $T^* = 0$ and $T^* = 3$ (Fig.~\ref{3.1-cv-Rg} (d), black
line), whereas that of block B increases by about 4 units (Fig.~\ref{3.1-cv-Rg}(f)). Despite this expansion
of block B, the total radius of gyration decreases by roughly one unit (Fig.~\ref{3.1-cv-Rg}(b), black line).
The expansion of block B only partially compensates for the shrinkage of block A, resulting
in an overall decrease in $R_g^2$. Fig.~\ref{3.1-Pit-N90} (b) shows the results of the knot localization map for the $3_1(45, 45)$. At low temperatures, i.e., in the
low-energy regime, the knot is restricted within the B block, as repulsive interactions in the A block favor its localization in this segment. In addition, the number of monomers in the
B block is sufficient to accommodate the knot, stabilizing its localization.

As shown in Fig.~\ref{3.1-cv-Rg}(d), the overall size of the copolymer ring is dominated by block A, whereas Fig.~\ref{3.1-cv-Rg}(e) shows that block B provides the main contribution to the total heat capacity.

For comparison, the behavior of a trefoil homopolymer under good and poor solvent conditions is shown in Fig.~\ref{3.1-cv-Rg}. In a good solvent (Homo($+$), Fig.~\ref{3.1-cv-Rg}(d)), the polymer adopts an expanded conformation at low temperatures. As the temperature increases, thermal fluctuations enhance configurational freedom, leading to a slight reduction in the radius of gyration and highlighting the increasing importance of entropic contributions to the free energy.
In a poor solvent (Homo($-$), Fig.~\ref{3.1-cv-Rg}(f)), the polymer adopts a compact conformation at low temperatures and gradually swells upon heating as thermal fluctuations overcome effective monomer--monomer attraction. The low-temperature peak in the heat capacity (Fig.~\ref{3.1-cv-Rg}(e)) originates from internal rearrangements within the compact globule without a significant change in size, while at higher temperatures the shoulder marks the onset of melting, accompanied by an increase in the radius of gyration (Fig.~\ref{3.1-cv-Rg}(f)).

\begin{figure}
  \begin{center}
    \includegraphics[width=0.48\textwidth]{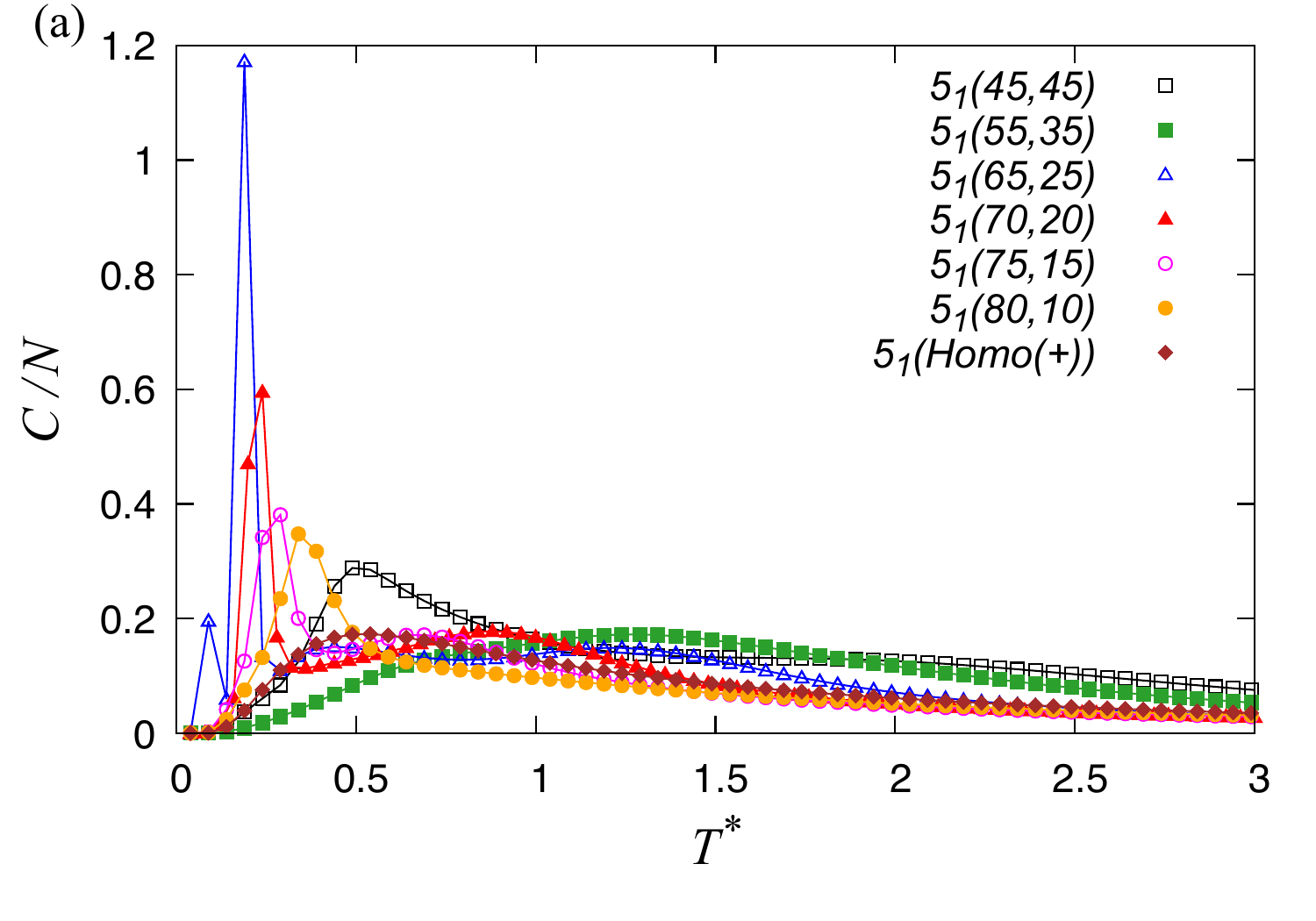}
      \includegraphics[width=0.48\textwidth]{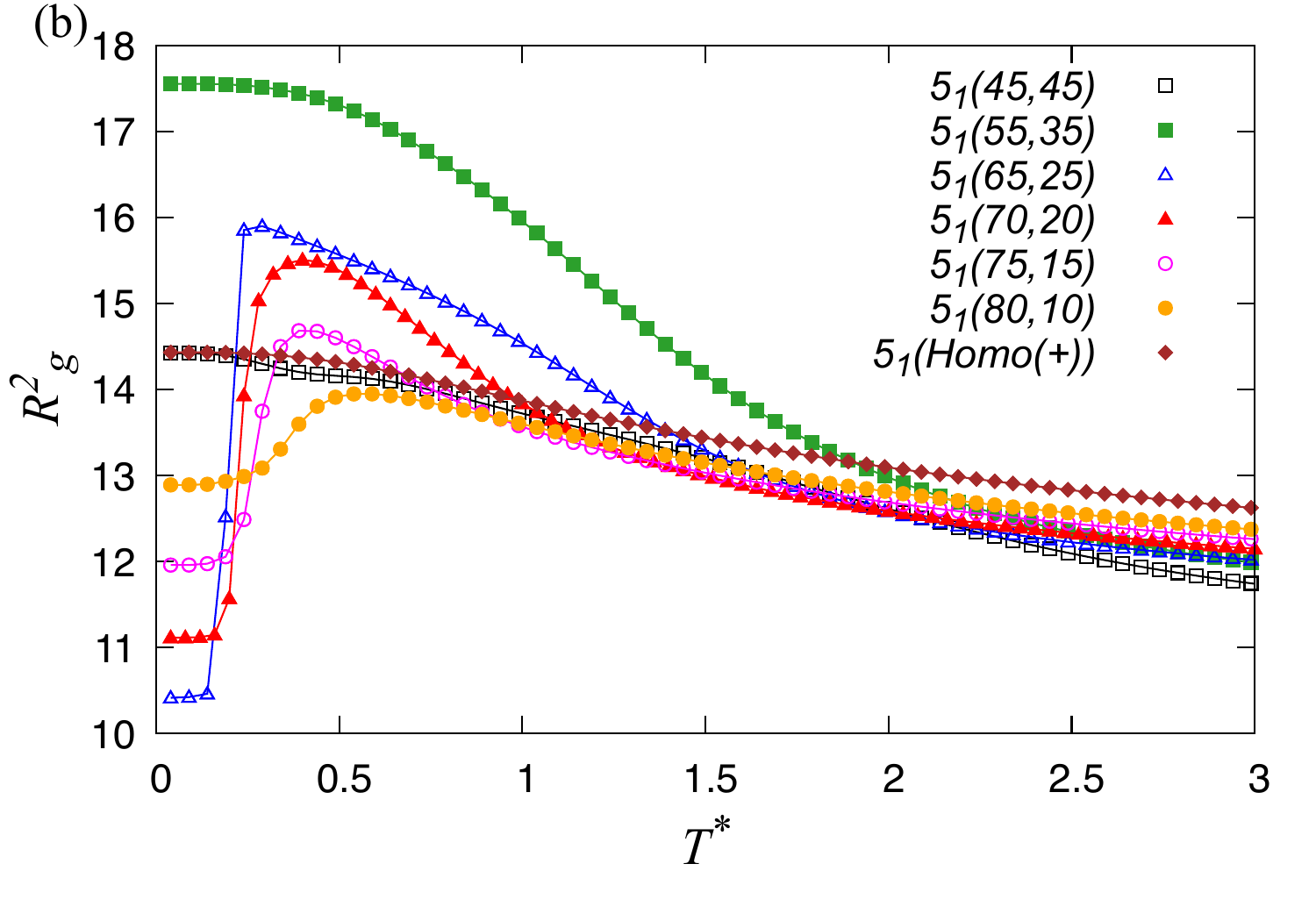}
 \caption{\justifying Thermodynamic and structural properties of knotted diblock copolymer rings with topology $5_1$ and total length $N=90$ for different monomer compositions. 
(a) The specific heat capacity $C/N$ (b) The mean-square radii of gyration $R_g^2$ as a function of temperature.}
\label{5.1-setupII} 
\label{5.1-gyr}
\end{center}
\end{figure}

\begin{figure}
\centering

\begin{subfigure}{0.48\textwidth}
\centering
\includegraphics[width=\linewidth]{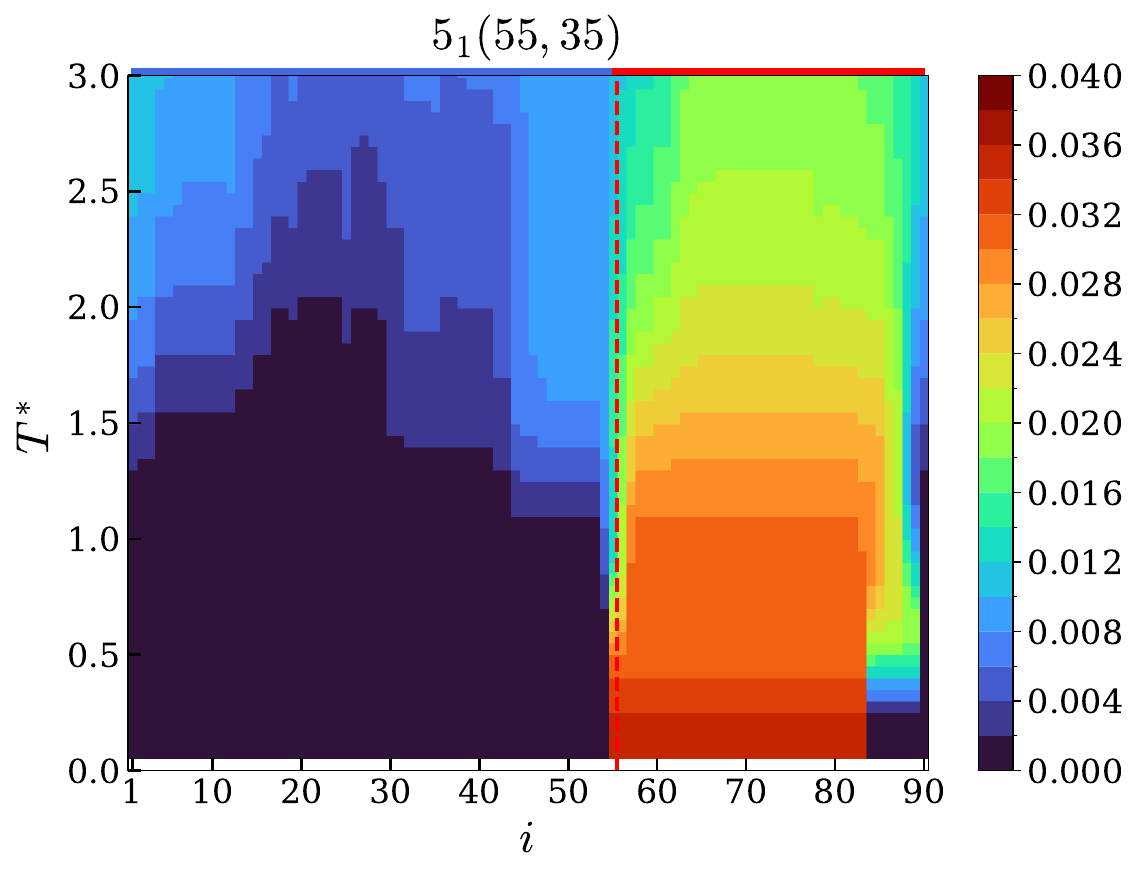}

\vspace{-18pt}
\centering (a)
\end{subfigure}
\hfill
\begin{subfigure}{0.48\textwidth}
\centering
\includegraphics[width=\linewidth]{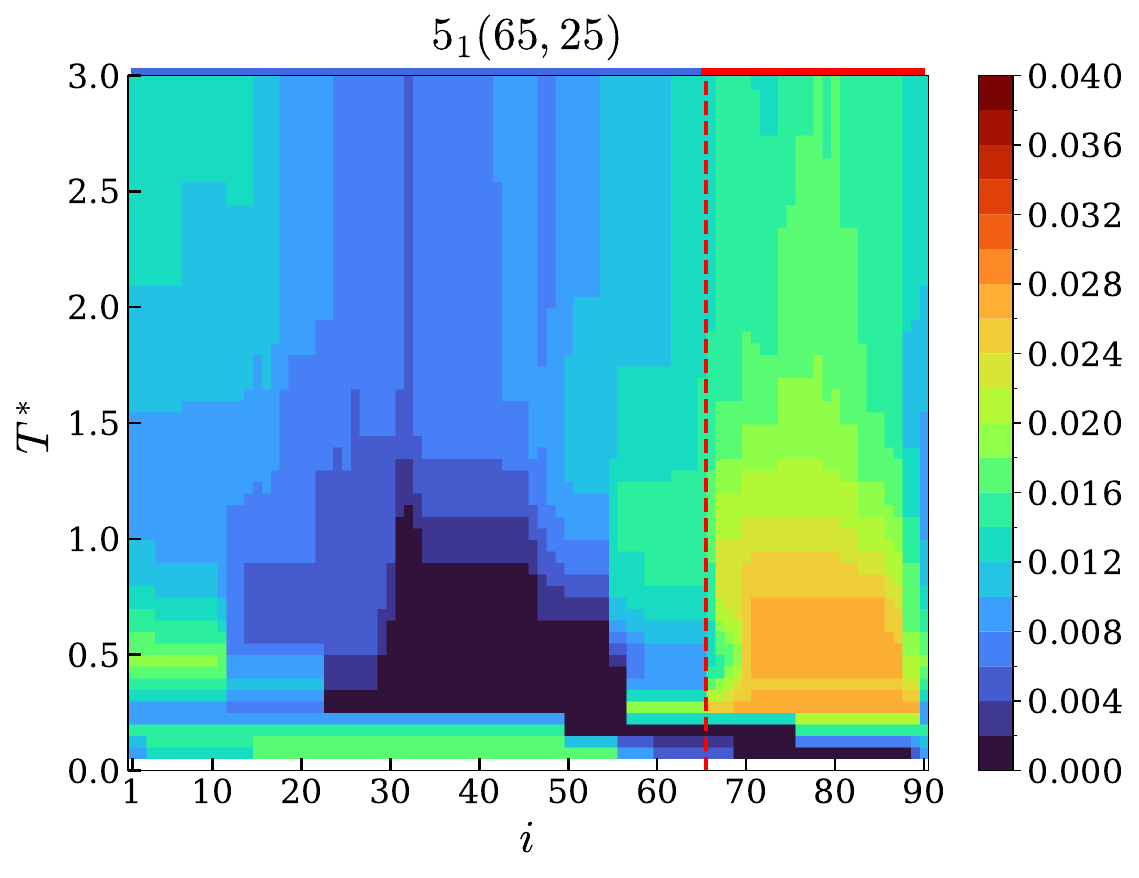}

\vspace{-18pt}
\centering (b)
\end{subfigure}
\caption{\justifying Knot localization maps for a pentafoil ($5_1$) diblock copolymer ring with monomer compositions (a) $5_1(55,35)$, (b) $5_1(65,25)$. The color scale represents the probability that monomer $i$ belongs to the minimal knotted region as a function of reduced temperature $T^*$ and monomer index $i$.}
\label{5.1-Pit-N90}
\end{figure}

\begin{figure}[t]
\centering
\begin{subfigure}{0.32\textwidth}
\centering
\begin{overpic}[width=\linewidth]{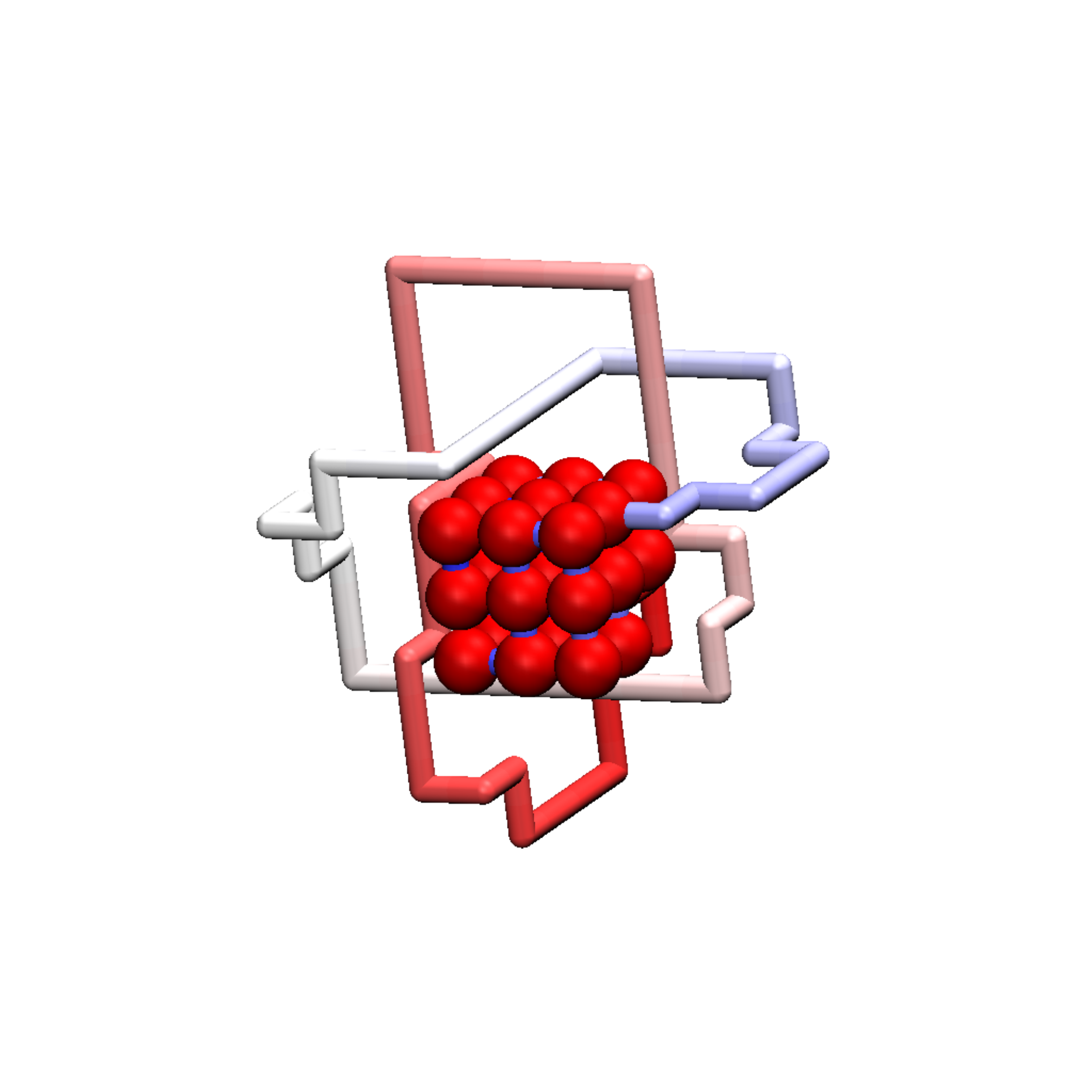}
\put(44,6){\normalsize (a)}
\end{overpic}
\end{subfigure}
\hfill
\begin{subfigure}{0.32\textwidth}
\centering
\begin{overpic}[width=\linewidth]{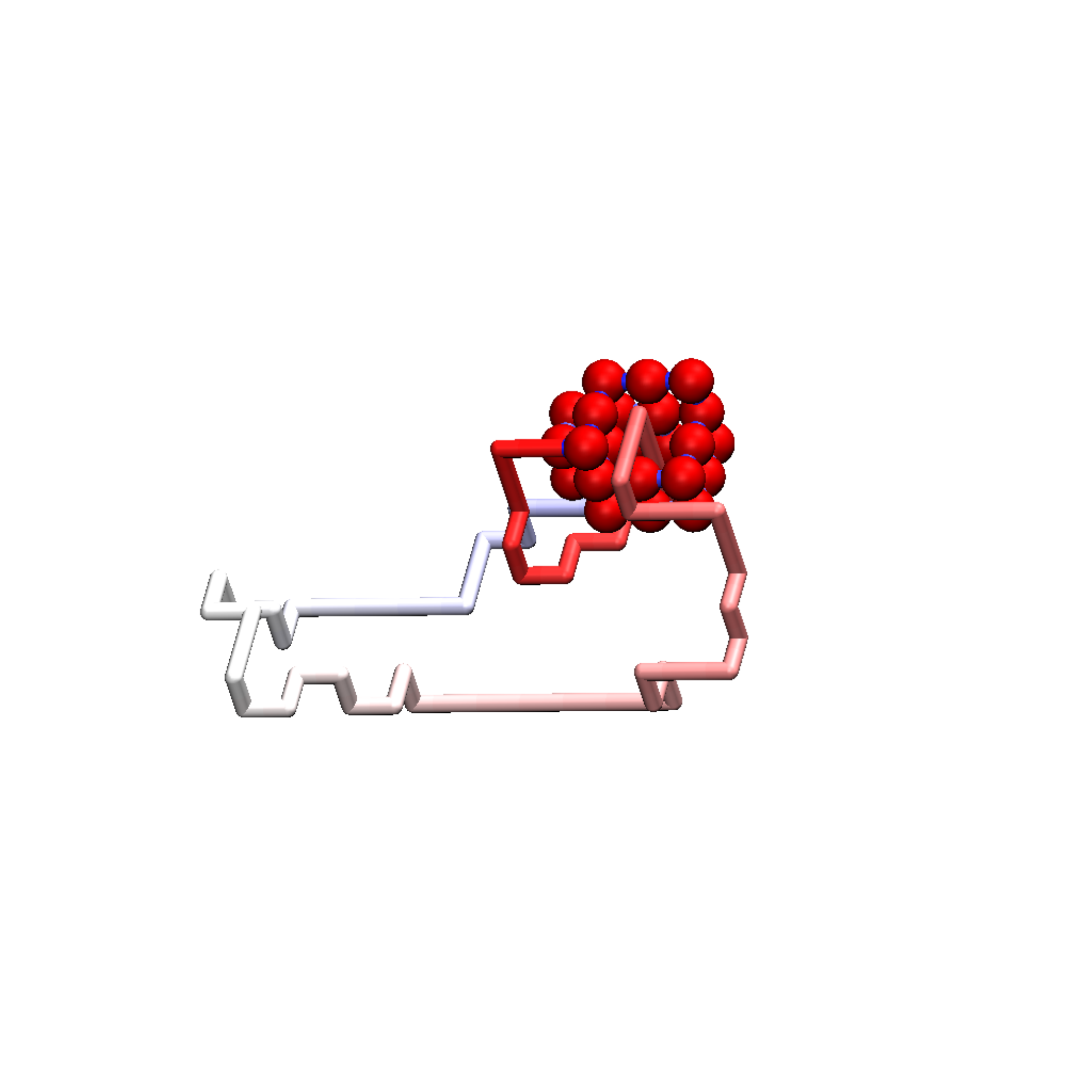}
\put(44,6){\normalsize(b)}
\end{overpic}
\end{subfigure}
\hfill
\begin{subfigure}{0.32\textwidth}
\centering
\begin{overpic}[width=\linewidth]{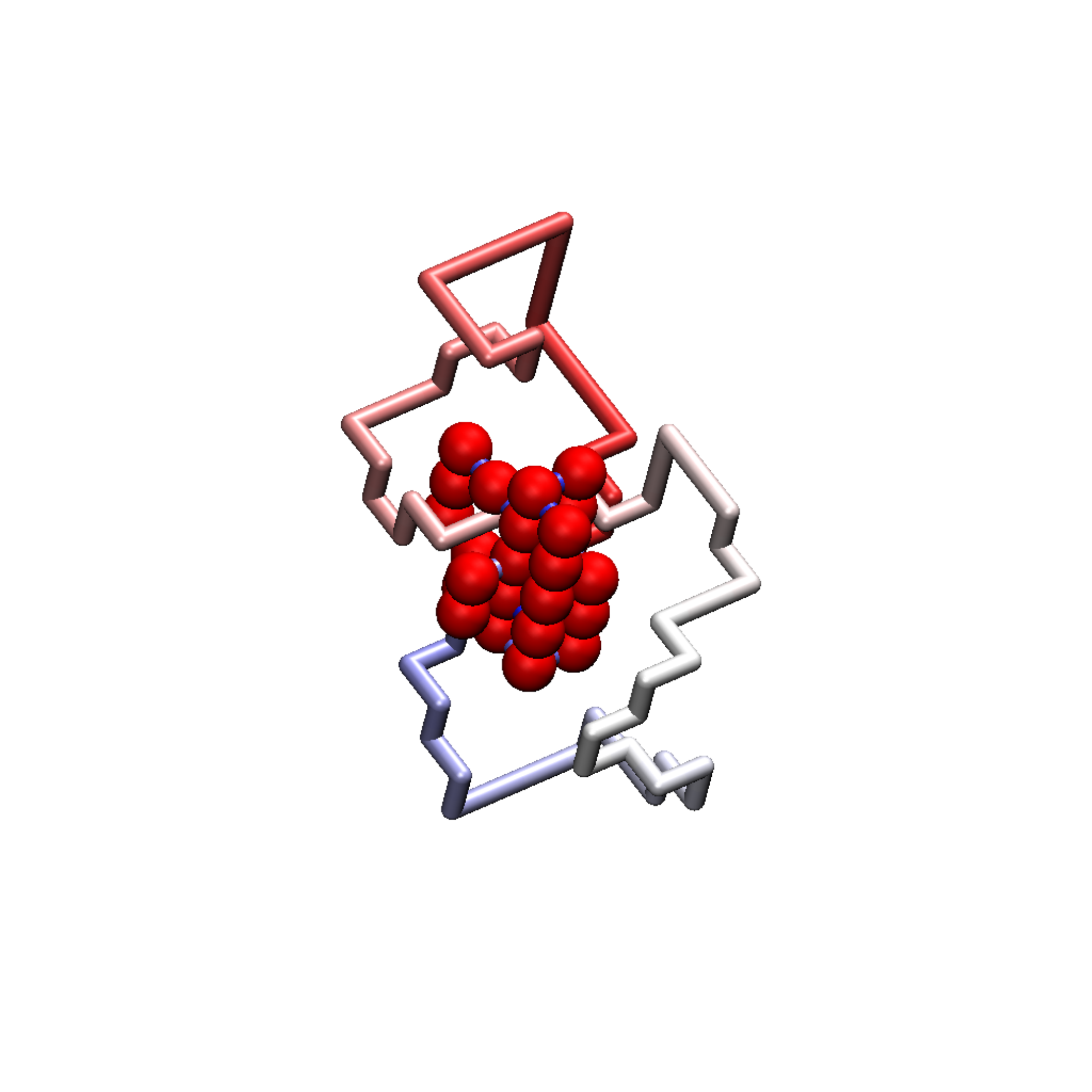}
\put(44,6){\normalsize(c)}
\end{overpic}
\end{subfigure}
\caption{\justifying Typical configurations of the copolymer ring with pentafoil topology and monomer composition $5_1(65,25)$ (a) the ground-state configuration at energy $E=-24$,
(b) a configuration with energy $E = -17$ at $T= 0.35$, and
(c) a configuration with energy $E = -10$ at $T=1$.
Red beads represent the attractive B monomers, while the A block is shown as lines.}
\label{5.1-conf}
\end{figure}

Next, we examine how the thermal and structural properties of the copolymer ring are
affected by the $5_1$ topology. Fig.~\ref{5.1-gyr} shows the heat capacity and the total radius of gyration
for $5_1$ knots with total length N = 90. The radius of gyration curves in panel  Fig.~\ref{5.1-gyr}(b) exhibit behavior similar to that of the $3_1$ knot;
however, the most pronounced reentrant-like behavior is observed for the longer B block
containing 25 monomers, i.e., $5_1(65, 25)$. This transition is characterized by a pronounced
peak in the heat capacity at \( T^* \simeq 0.2 \) (Fig.~\ref{5.1-gyr}(a), blue line). For monomer compositions with fewer than 25 B-type monomers, the reentrant-like behavior (nonmonotonic behavior in Fig.~\ref{5.1-gyr}(b)) is still present but becomes considerably weaker as the B-block length decreases. In contrast, for the composition $5_1(55, 35)$, the copolymer
exhibits a larger overall size compared to the other monomer distributions. This behavior
can be related to the minimal length required to accommodate the $5_1$ knot, which is 34
monomers on a cubic lattice. In this case, the knot localizes within the B block, allowing the A block to extend more freely and resulting in an increased overall size. At high
temperatures, the mean-square radii of gyration for all diblock copolymer rings converge
to a common value, indicating that thermal fluctuations dominate over monomer–monomer
interactions and the system behaves effectively as a homopolymer.
Fig.~\ref{5.1-Pit-N90}(a) shows the knotted region for the case $5_1(55, 35)$. The knot is strongly localized
within the B block at low temperatures \( T^{*} \lesssim 1.5 \), as indicated by the red region. At higher
temperatures, the localization weakens (green region), and the knot progressively delocalizes
into the A block (light blue regions). The color map in the panel (b) of Fig.~\ref{5.1-Pit-N90} confirms that in the monomer composition $5_1(65,25)$ the knot is not localized within the B block until the temperature of approximately $T^*=0.2$ is reached, corresponding more or less to the high peak observed in the plot of the specific heat capacity. As it is possible to note in Fig.~\ref{5.1-Pit-N90}(b), the localization of the knot is only partial (yellow and orange colors), because the size of the B block is 25 monomers only, while the minimal size of the $5_1$ knot on the simple cubic lattice is 34. Typical configurations of copolymer $5_1(65, 25)$ are shown in Fig.~\ref{5.1-conf}. In panel (a), corresponding to the lowest investigated temperature, the knot is not localized within the B block as discussed previously. In panel (b), the knot is localized at $T^*=0.35$, and a partial localization is still present at $T^*=1$.
\subsection{ THERMAL PROPERTIES OF LONG KNOTTED DIBLOCK COPOLYMER RINGS ( N=200 )}~\label{long}
We extend our simulations to diblock copolymer rings with total length
$N = 200$ to examine how the B-block length required to induce
reentrant-like behavior depends on the overall chain length.

\begin{figure}
  \begin{center}
    \includegraphics[width=0.48\textwidth]{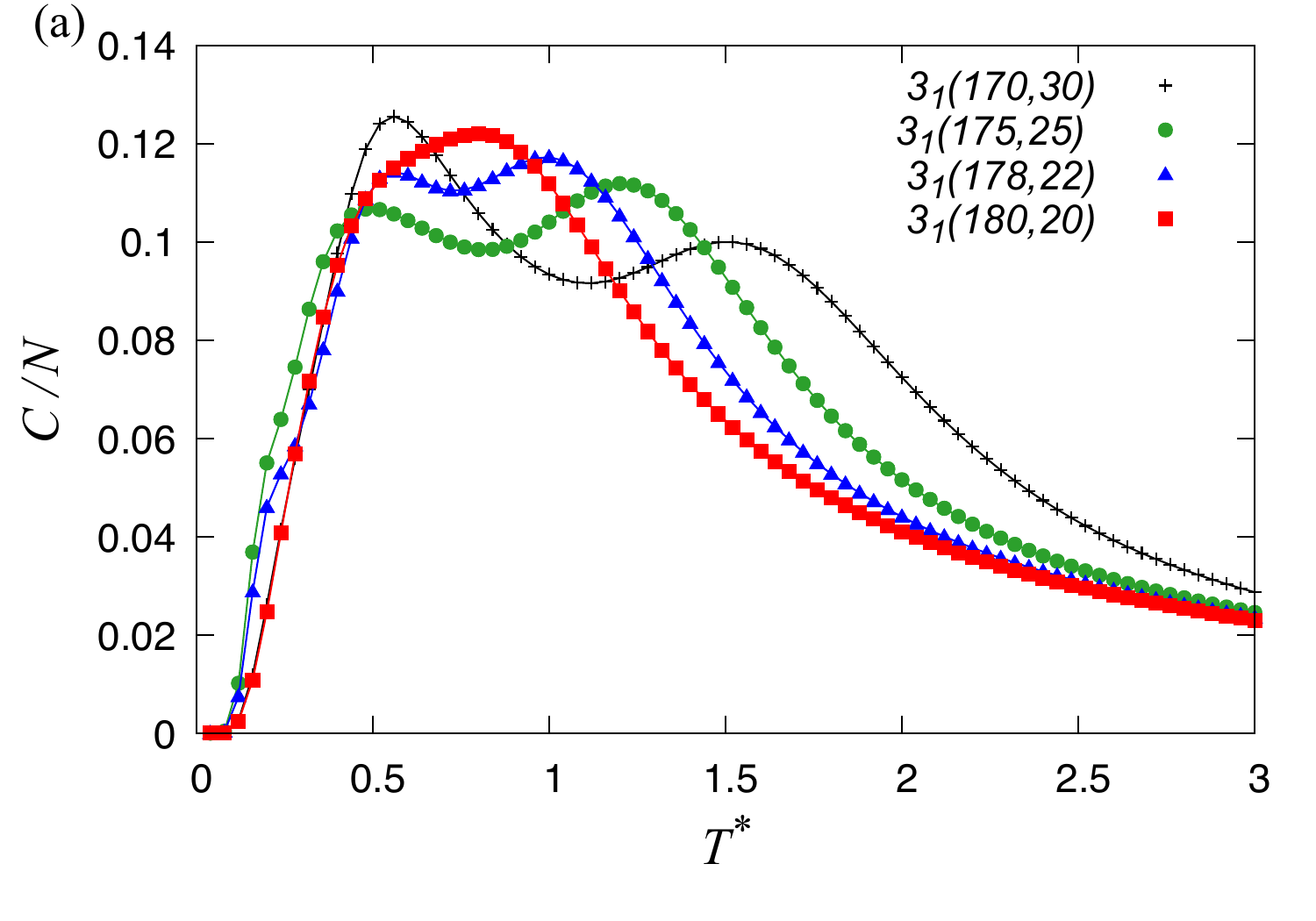}
    \includegraphics[width=0.48\textwidth]{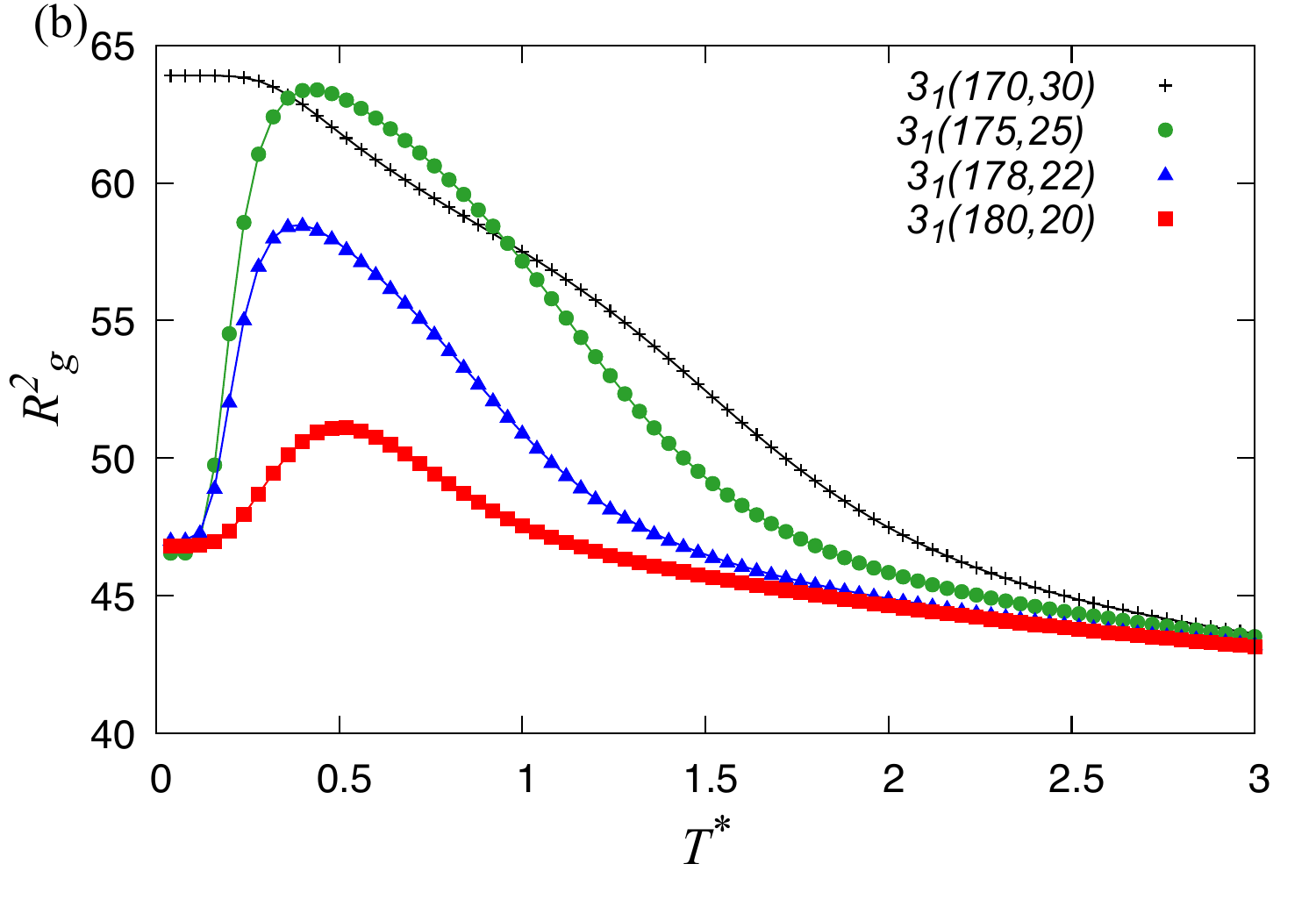}
   \caption{ \justifying (a) Specific heat capacities $C/N$ as a function of reduced temperature $T^*$ for trefoil ($3_1$) knotted diblock copolymer rings with $N=200$ and different monomer compositions.
(b) Mean-square radii of gyration $R_g^2$ versus $T^*$. }
\label{3.1-N200} 
\end{center}
\end{figure}

Fig.~\ref{3.1-N200} shows the heat capacity and the mean-square radius of gyration for diblock copolymers with trefoil topology $3_1$ and different monomer compositions. As shown in Fig.~\ref{3.1-N200}(b), at low temperatures the system $3_1(170,30)$ exhibits the largest overall size, whereas the most pronounced reentrant-like behavior occurs for the composition $3_1(175,25)$ (green line). This behavior emerges for B-block lengths of
approximately 25 monomers and becomes gradually weaker as the B-block
length is reduced toward 20 monomers.

\begin{figure}
  \begin{center}
    \includegraphics[width=0.48\textwidth]{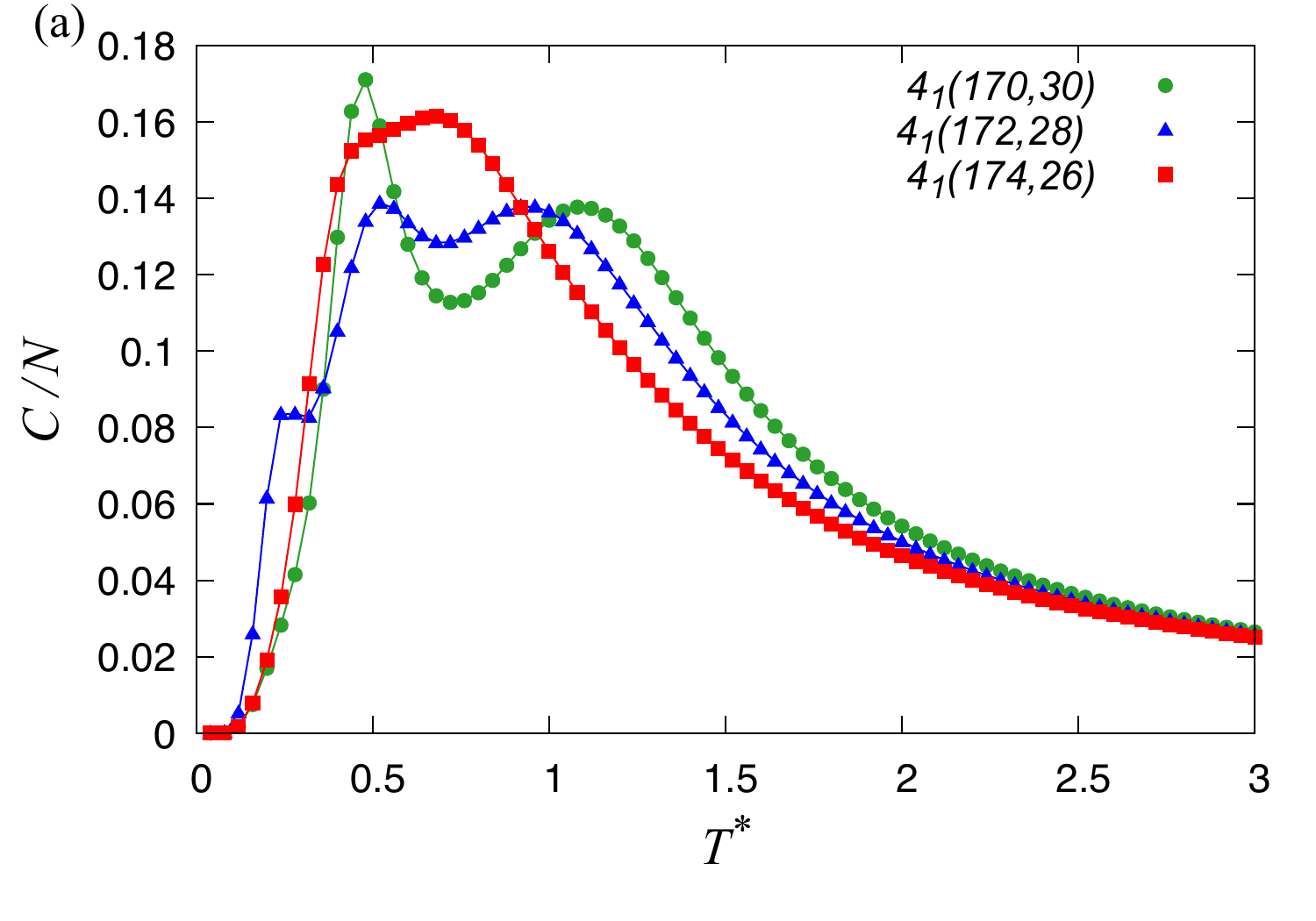}
    \includegraphics[width=0.48\textwidth]{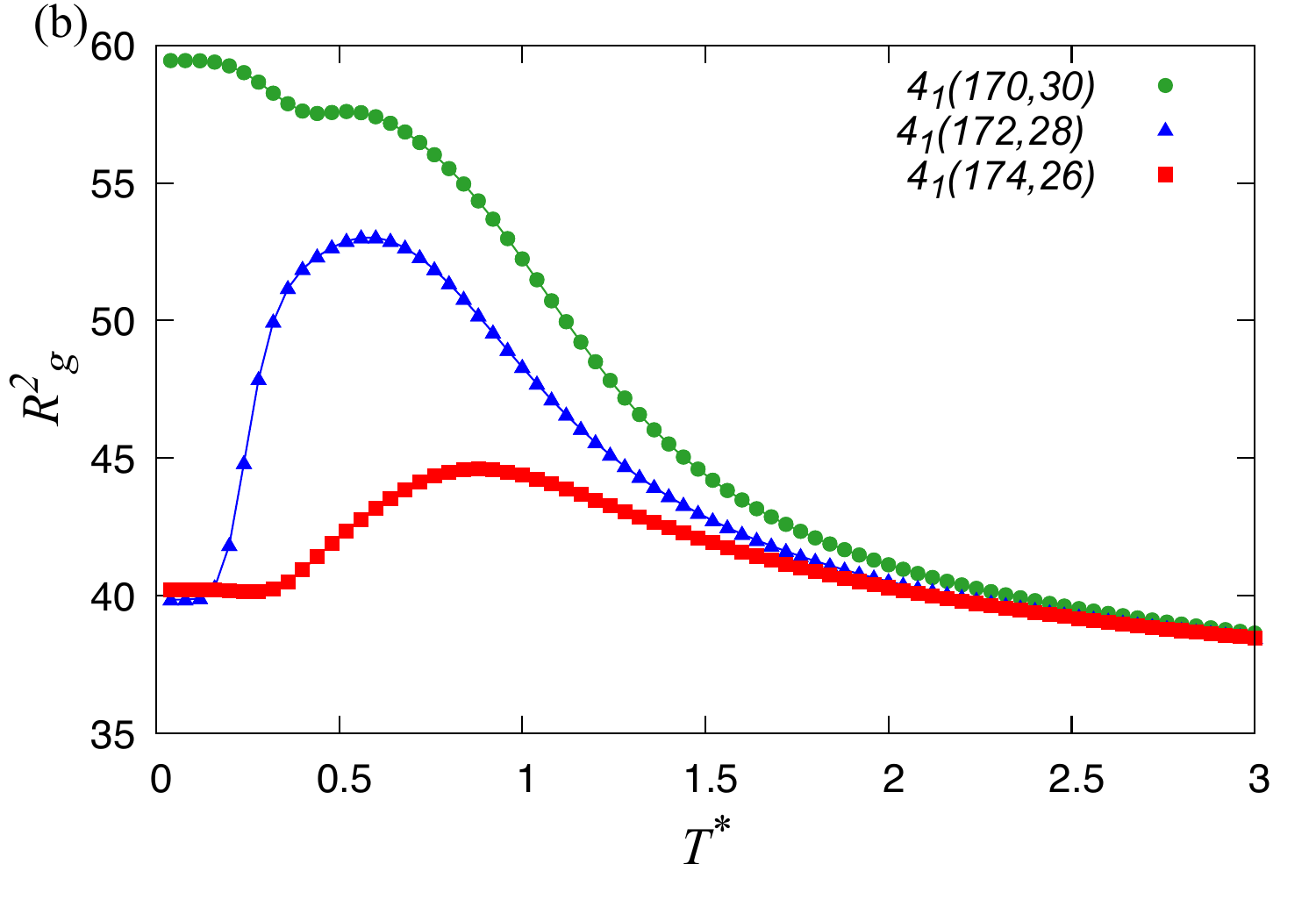}
   \caption{\justifying (a) Specific heat capacities $C/N$ as a function of reduced temperature $T^*$ for figure-eight ($4_1$) knotted diblock copolymer rings with $N=200$ and different monomer compositions.
(b) Mean-square radii of gyration $R_g^2$ versus $T^*$. }
\label{4.1-N200} 
\end{center}
\end{figure}

\begin{figure}
  \begin{center}
    \includegraphics[width=0.48\textwidth]{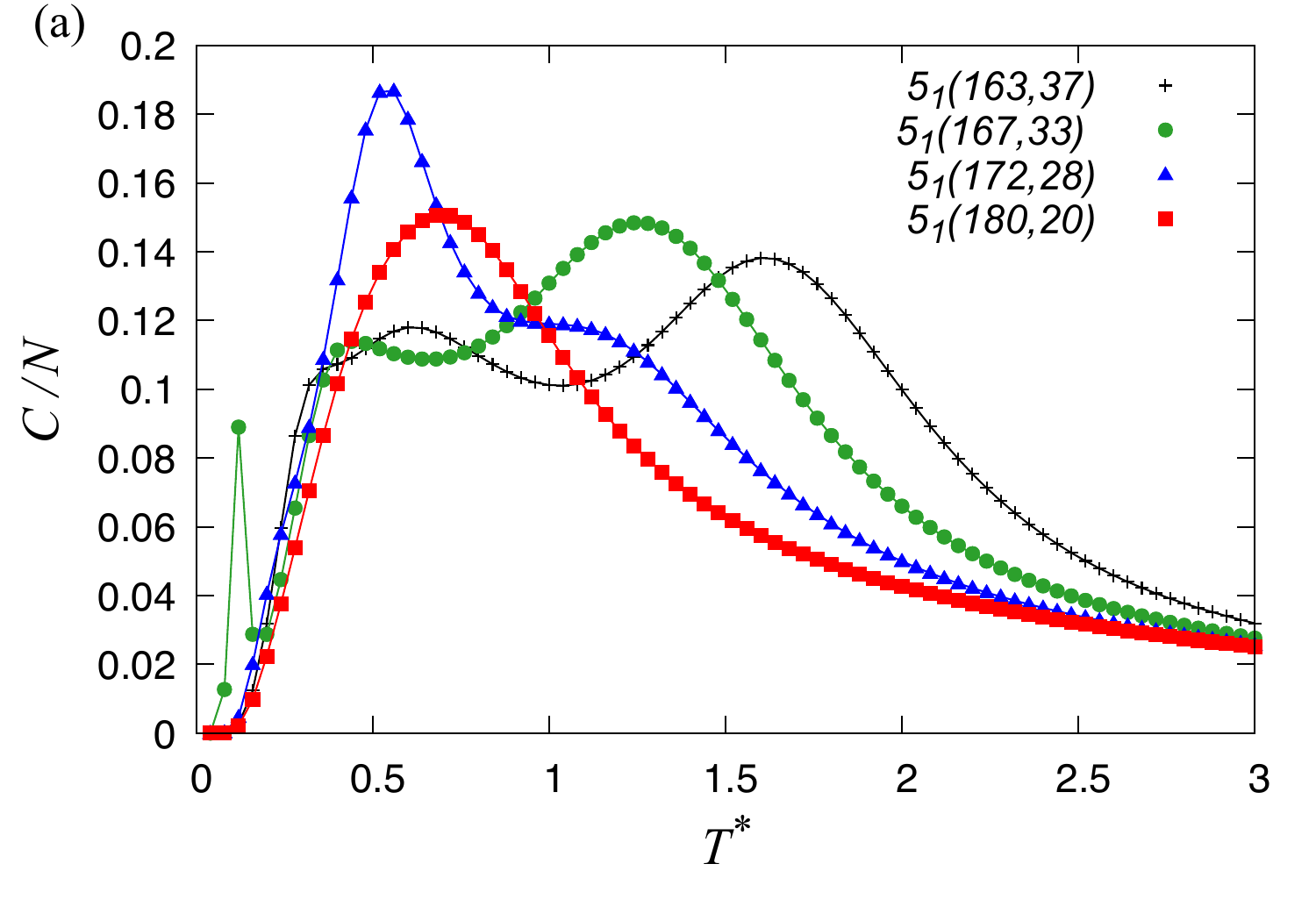}
    \includegraphics[width=0.48\textwidth]{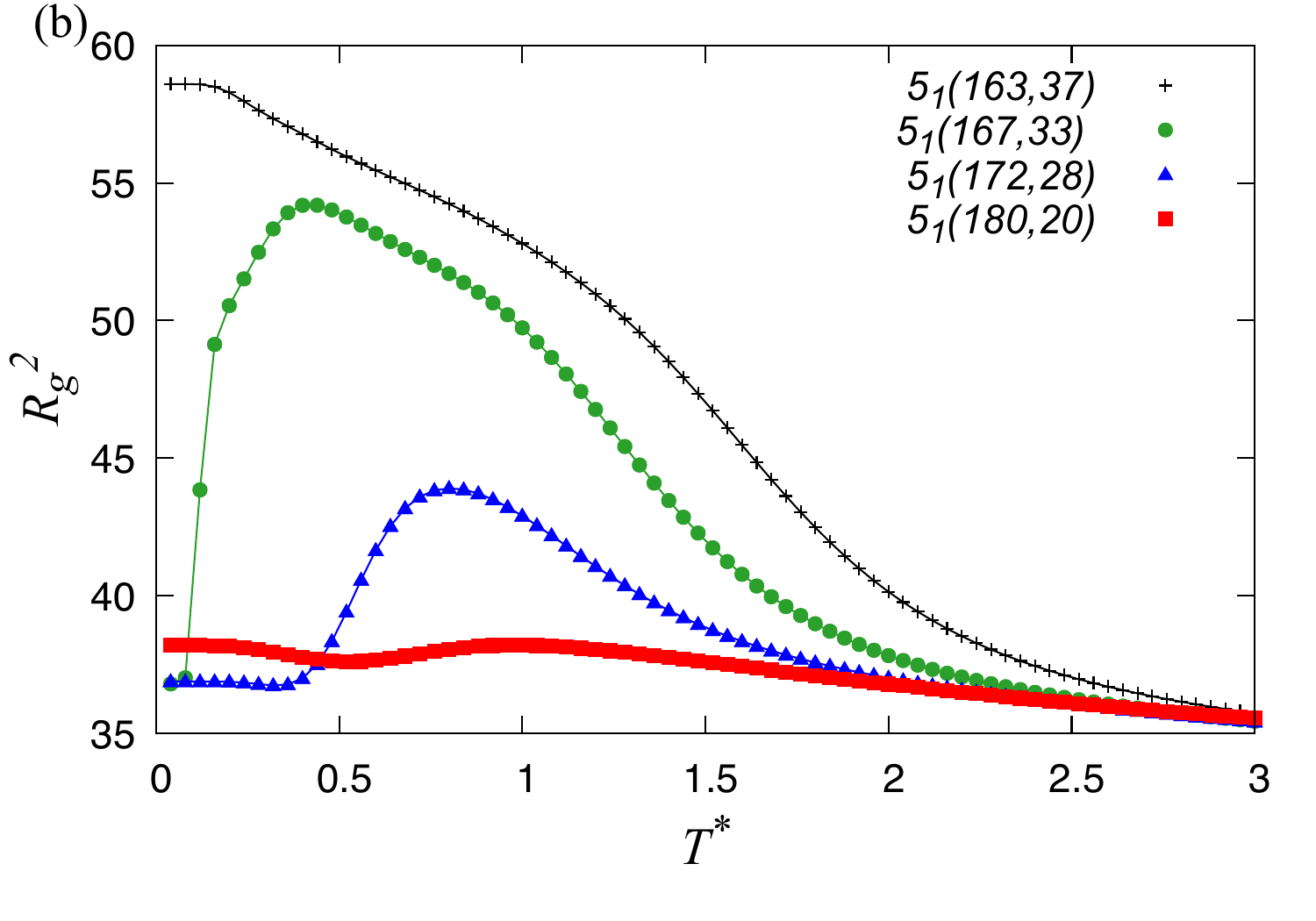}
     \includegraphics[width=0.48\textwidth]{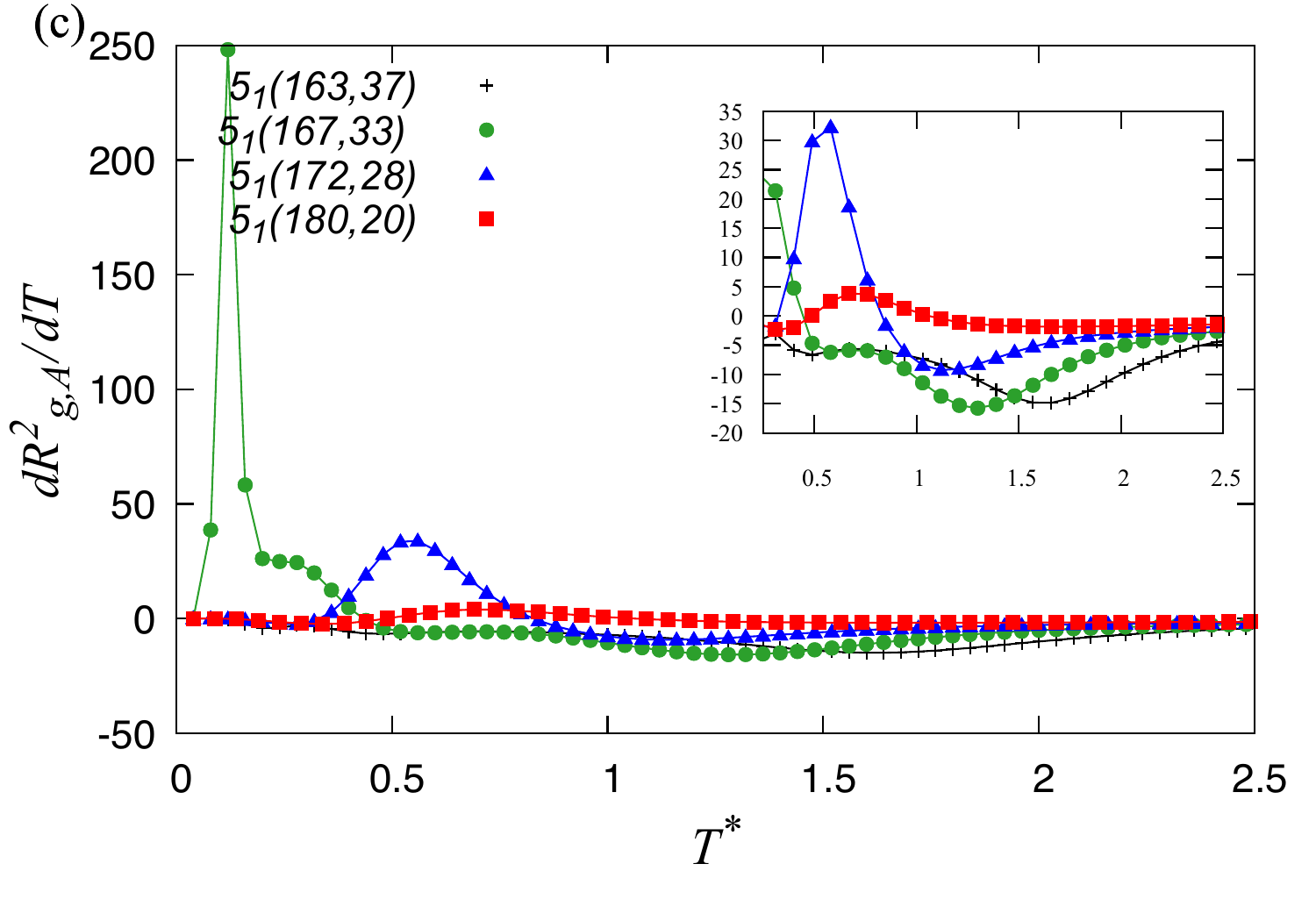}
    \includegraphics[width=0.48\textwidth]{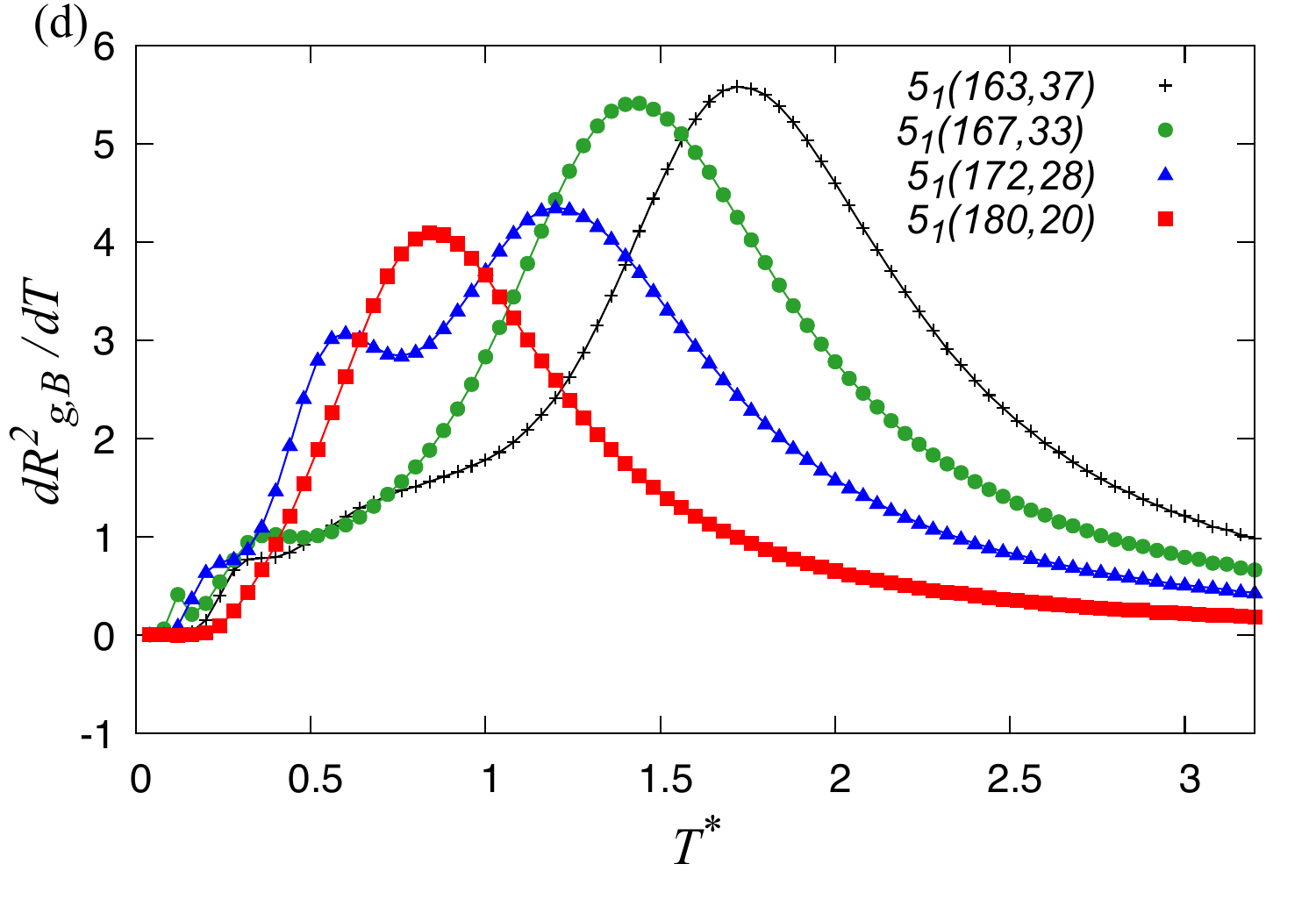}
\caption{\justifying Thermodynamic and structural properties of knotted diblock
copolymer rings with the $5_1$ knot and total length $N=200$ for different
monomer compositions.
(a) Specific heat capacity per monomer $C/N$ vs reduced temperature $T^*$.
(b) Mean-square radius of gyration $R_g^2$ vs $T^*$. 
(c) Temperature derivative of the mean-square radius of gyration of the A block, $dR_{g,A}^2/dT^*$ (inset: enlarged vertical scale). 
(d) Temperature derivative of the mean-square radius of gyration of the B block, $dR_{g,B}^2/dT^*$.}
\label{5.1-N200} 
\end{center}
\end{figure}

A similar nonmonotonic response is observed for the figure-eight knot $(4_1)$ with total length $N = 200$ when the B block contains approximately 28 monomers, as shown by the blue curve in Fig.~\ref{4.1-N200}(b). All monomer compositions display a main peak in the specific heat capacity around $T^* \sim 0.5$ (Fig.~\ref{4.1-N200}(a)).
However, for the composition $4_1(172,28)$ an additional low-temperature
shoulder appears. This feature can be attributed to the largest nonmonotonic variation of the mean-square radius of gyration.
For the pentafoil ($5_1$) knot, we performed the simulations for various monomer compositions. Among the cases considered, $5_1(163,37)$ exhibits the largest overall radius of gyration, whereas $5_1(167,33)$ is significantly more compact at $T^* = 0.05$, with a radius of gyration nearly half that of $5_1(163,37)$ (Fig.~\ref{5.1-N200}(b)).
The specific heat capacity of $5_1(167,33)$ displays multiple peaks (Fig.~\ref{5.1-N200}(a)), indicating the presence of several structural transitions. In particular, three distinct peaks are observed, while $5_1(180,20)$ exhibits only a single peak, suggesting more complex conformational rearrangements in the former system.

\begin{figure}
\centering

\begin{subfigure}{0.48\textwidth}
\centering
\includegraphics[width=\linewidth]{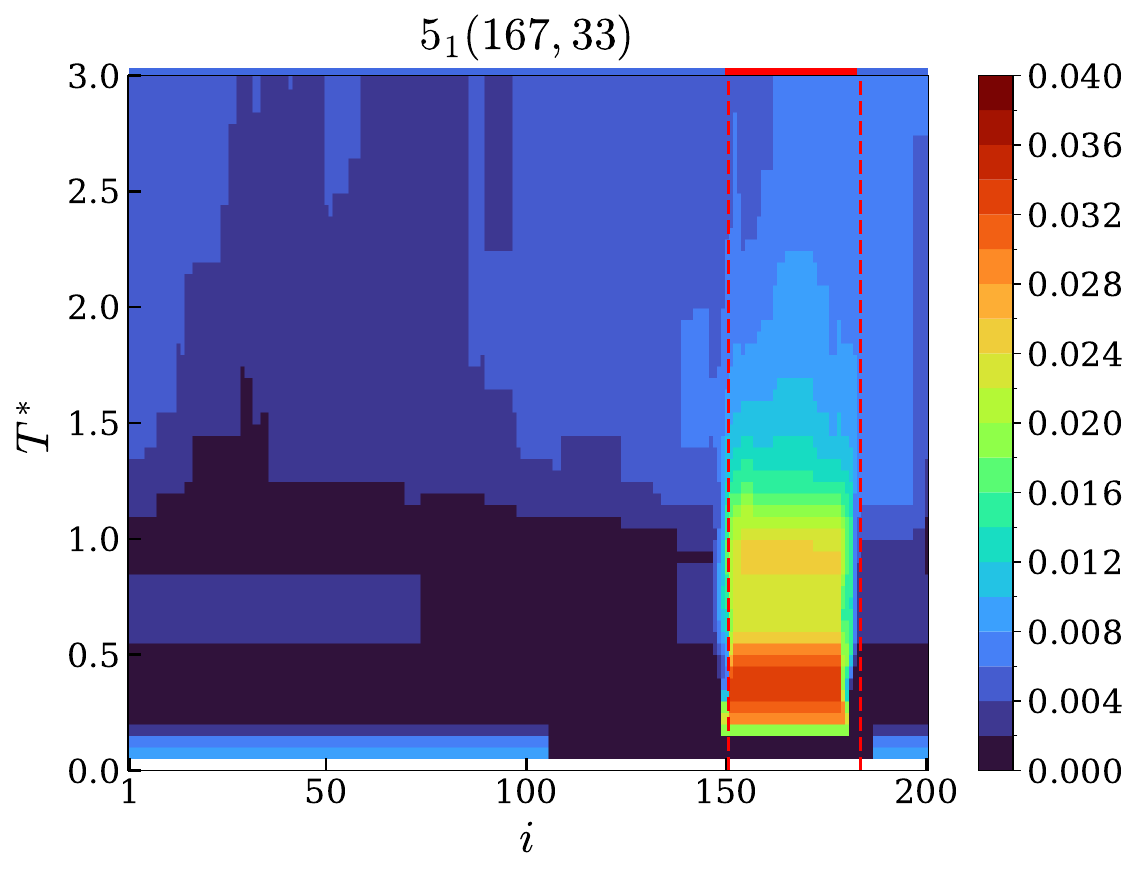}

\vspace{-18pt}
\centering \small (a)
\end{subfigure}
\hfill
\begin{subfigure}{0.48\textwidth}
\centering
\includegraphics[width=\linewidth]{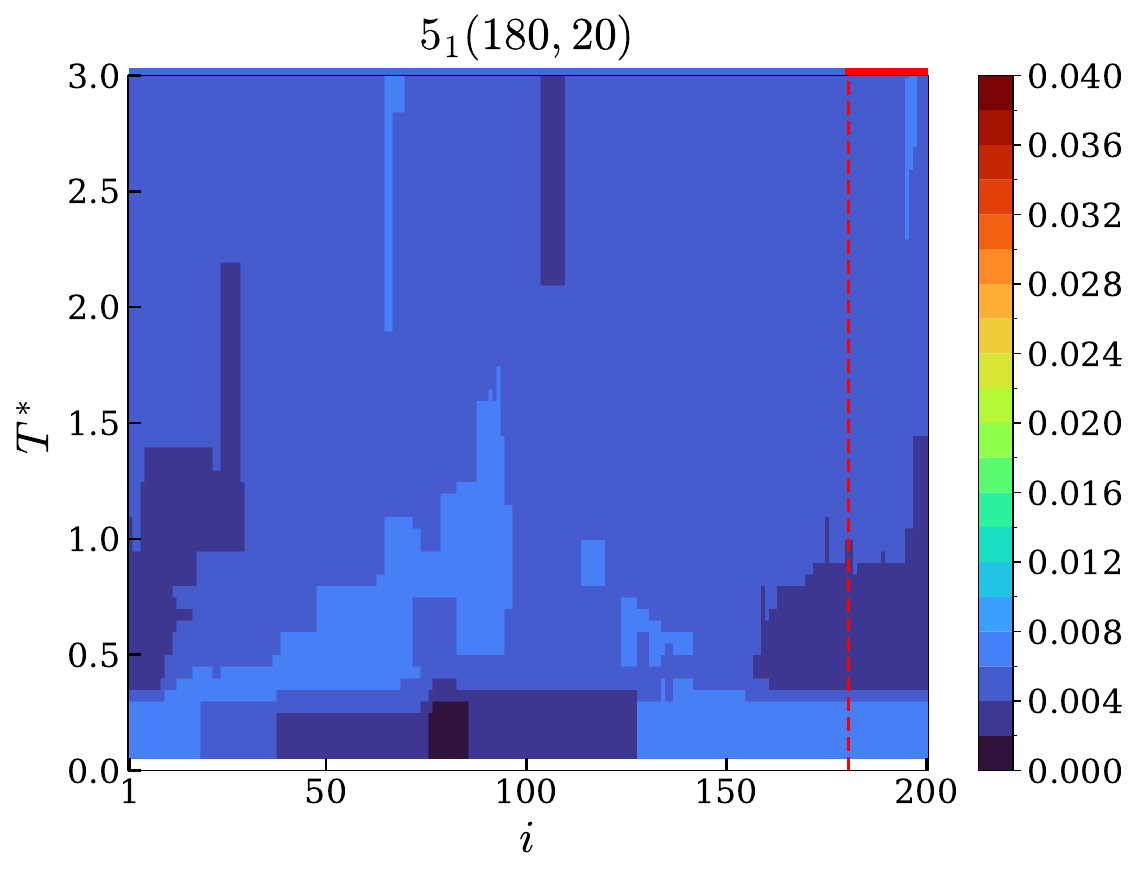}

\vspace{-18pt}
\centering \small (b)
\end{subfigure}

\caption{\justifying Representative color maps showing the likelihood that a bead belongs to the knotted segment as a function of temperature $T^*$ and monomer index $i$ for diblock copolymer rings with knot $5_1$ and $N=200$: (a) $5_1(167,33)$ and (b) $5_1(180,20)$. The vertical line denotes the $A$--$B$ interface between the A and B blocks, while the top line shows the block regions.}

\label{5.1-localization-map}
\end{figure}

To further disentangle the block-specific contributions, we analyze the temperature derivatives $dR^2_{g,A}/dT^*$ and $dR^2_{g,B}/dT^*$, which allow us to identify the characteristic temperatures of structural changes in each block. In Fig.~\ref{5.1-N200}(c), the pronounced peak in $dR^2_{g,A}/dT^*$ for $5_1(167,33)$ indicates a sharp conformational transition of the A block occurring over a narrow temperature interval. This behavior arises from a rapid rearrangement triggered by the abrupt localization of the knot within the B block, which relaxes topological constraints on the A block and enables it to adopt a more extended configuration.
The knot localization map for $5_1(167,33)$, Fig.~\ref{5.1-localization-map}(a), exhibits red and green regions, indicating the localization of the knot over the low temperature range $0.1 \lesssim T^* \lesssim 1$.  As the temperature increases and the B block melts, the knot gradually delocalizes and spreads over a larger portion of the chain.

For comparison, we consider here also the knot localization in a ring with a smaller B-size, namely the ring $5_1(180, 20)$ shown in Fig.~\ref{5.1-localization-map}(b). 
As illustrated, the diblock copolymer with monomer composition $5_1(180,20)$ does not exhibit strong knot localization, in contrast to $5_1(167,33)$, where the knot is strongly localized, as indicated in Fig.~\ref{5.1-localization-map}(a) by a region of lighter colors centered around the B block (monomers 150--183).
Instead,  in $5_1(180, 20)$, the knot is more delocalized and tends to spread over
the entire ring, as shown by the widespread area with blue colors. As a consequence, in the latter case, the reentrant-like conformational behavior
disappears. 

For the case $5_1(172,28)$, two extrema appear in $dR_{g,A}^2/dT^*$ (Fig.~\ref{5.1-N200}(c), blue curve): a maximum near $T^* \simeq 0.5$, marking expansion of the A block and coinciding with an increase in $dR_{g,B}^2/dT^*$, and a minimum around $T^* \simeq 1.2$, indicating shrinkage of the A block and relocalization of the knot along the chain; this is accompanied by melting of the B block in Fig.~\ref{5.1-N200}(d).
 
Fig.~\ref{5.1-N200}(d) shows a small bump at very low temperature for monomer compositions $5_1(167,33)$ corresponding to the rearrangement of the B block following the expansion during the knot relocation as shown in Fig.~\ref{5.1-N200}(c). The temperatures of the main peaks in Fig.~\ref{5.1-N200}(d) increase with the size of the B block. This reflects the fact that a larger B block stabilizes the collapsed state of the B monomers.
 
To highlight the distinct behavior of the $5_1(167, 33)$ composition, it is instructive to compare its reentrant-like behavior in the plots of the specific heat capacity with the analogous behaviors occurring in the simpler knots $3_1$ and $4_1$ in the case of monomer compositions $3_1(175,25)$ and $4_1(172,28)$, respectively.
The comparison will also be extended to the shorter chain cases $5_1(65, 25)$ and $5_1(55, 35)$.
In general, we have seen that for different knot topologies and ring sizes, there is a monomer composition that corresponds to the maximal reentrant behavior, which is characterized by the fact that the growth of the gyration radius is the most pronounced.
In some cases, like for instance $5_1(65,25)$ and $5_1(167,33)$, the specific heat capacity exhibits three distinct peaks.
The peak at the lowest temperature (less than $ T^* \sim 0.5$) is due to knot relocation. The peak at the highest temperature can be attributed to the melting of the B block.
The intermediate peak is due to a rearrangement of the B-monomers following a small relaxation of the localization of the knot in the B block with increasing temperature. The probable reason for this rearrangement is the passage from compact-ordered conformations to compact-disordered conformations, as discussed in the paper of~\cite{taylor2009phase}, where these conformations have been called crystallites and collapsed globules, respectively.

For the maximally reentrant monomer compositions $3_1(175, 25)$ and $4_1(172, 28)$, the low-temperature rearrangement of the B
 block and the relocation of the knot occur over a broader temperature range, so that the total effect of these two transformations is a single
 broad peak in the specific heat capacity around $T^* \sim 0.5 $, see Figs.~\ref{3.1-N200}(a) and ~\ref{4.1-N200}(a). In contrast, for the $5_1(167, 33)$ case, the peak associated with
 knot relocation emerges within a narrow temperature interval at lower temperatures.

The difference observed for the pentafoil knot between the cases $N=90$ with composition $5_1(55,35)$ and $N=200$ with $5_1(167,33)$, despite the comparable length of the B block (see Figs.~\ref{5.1-gyr}(a) and \ref{5.1-N200}(a)), suggests that the conformational response is not determined solely by the absolute B-block length required to accommodate the knot, but also by its relative monomer fraction within the ring. While for $N=90$ and $N_B=35$ the knot can be fully localized within the B block, resulting in a smooth and gradual shrinkage upon heating, for $N=200$ and $N_B=33$ the localization becomes frustrated, leading to a low-temperature localization transition and a pronounced peak in the heat capacity. This explains why small variations in $N_B$ can lead to qualitative changes in the conformational behavior of the knotted copolymers with different sizes.

\begin{figure}
  \begin{center}
   \includegraphics[width=0.48\textwidth]{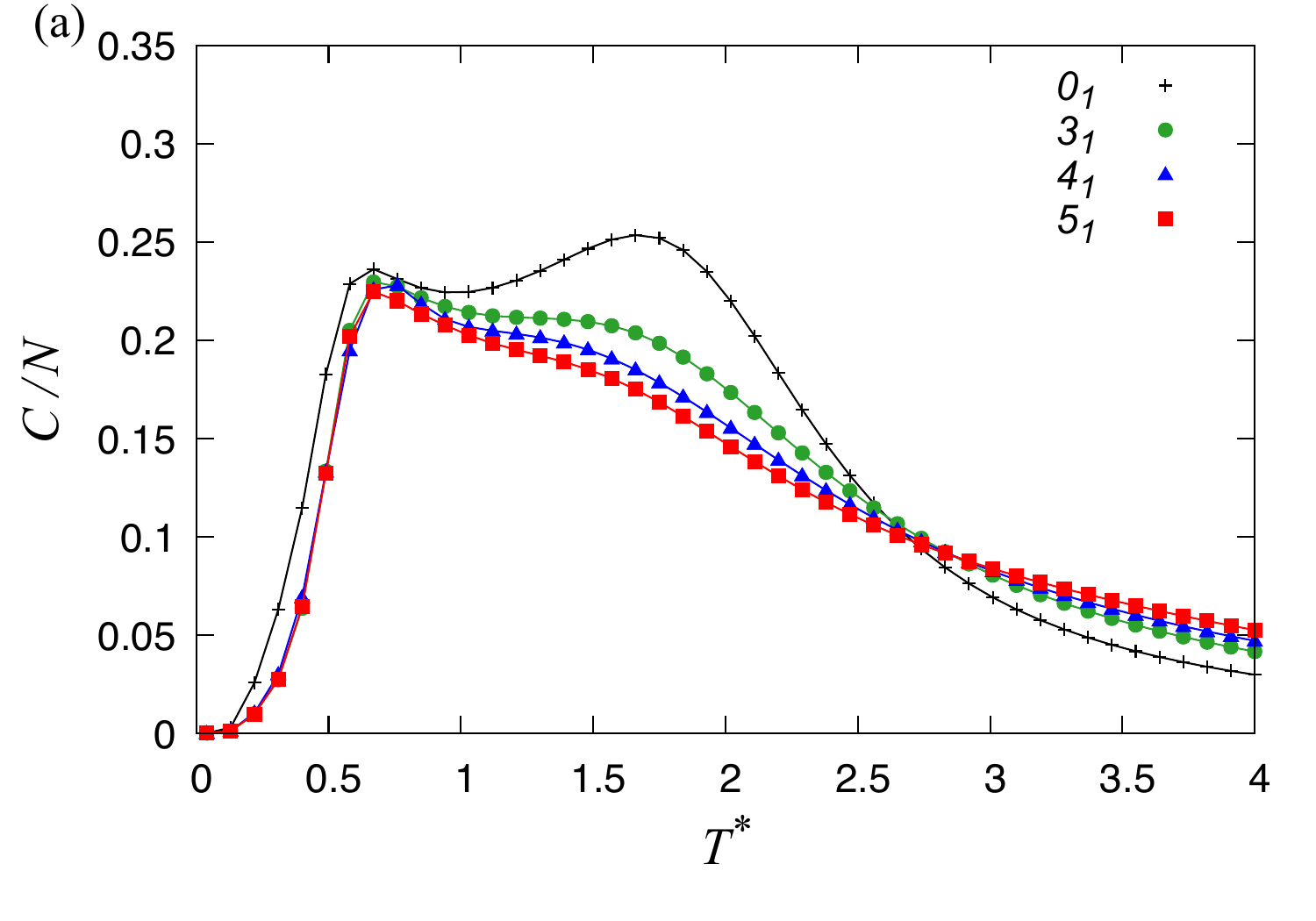}   
      \includegraphics[width=0.48\textwidth]{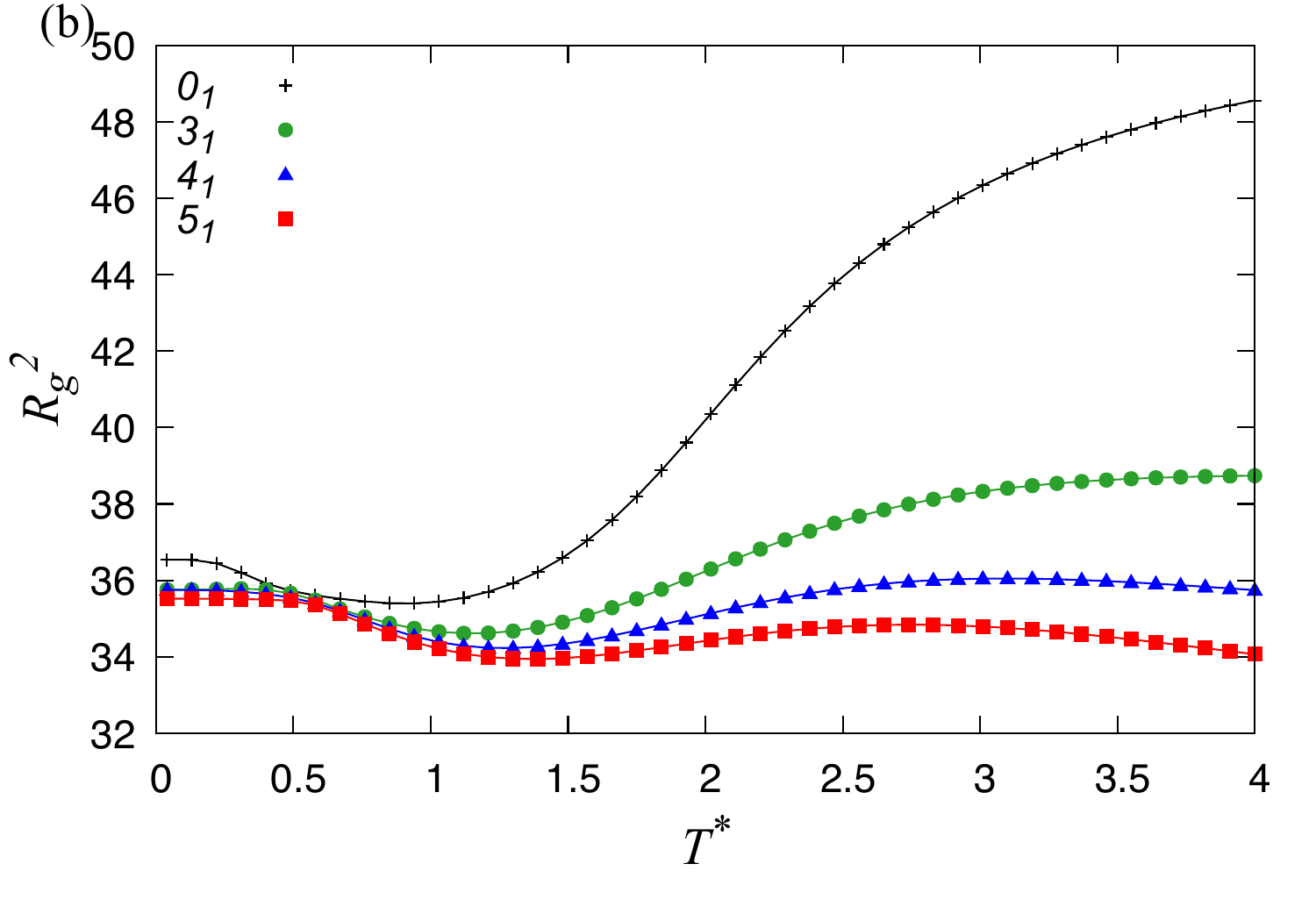}
     \includegraphics[width=0.48\textwidth]{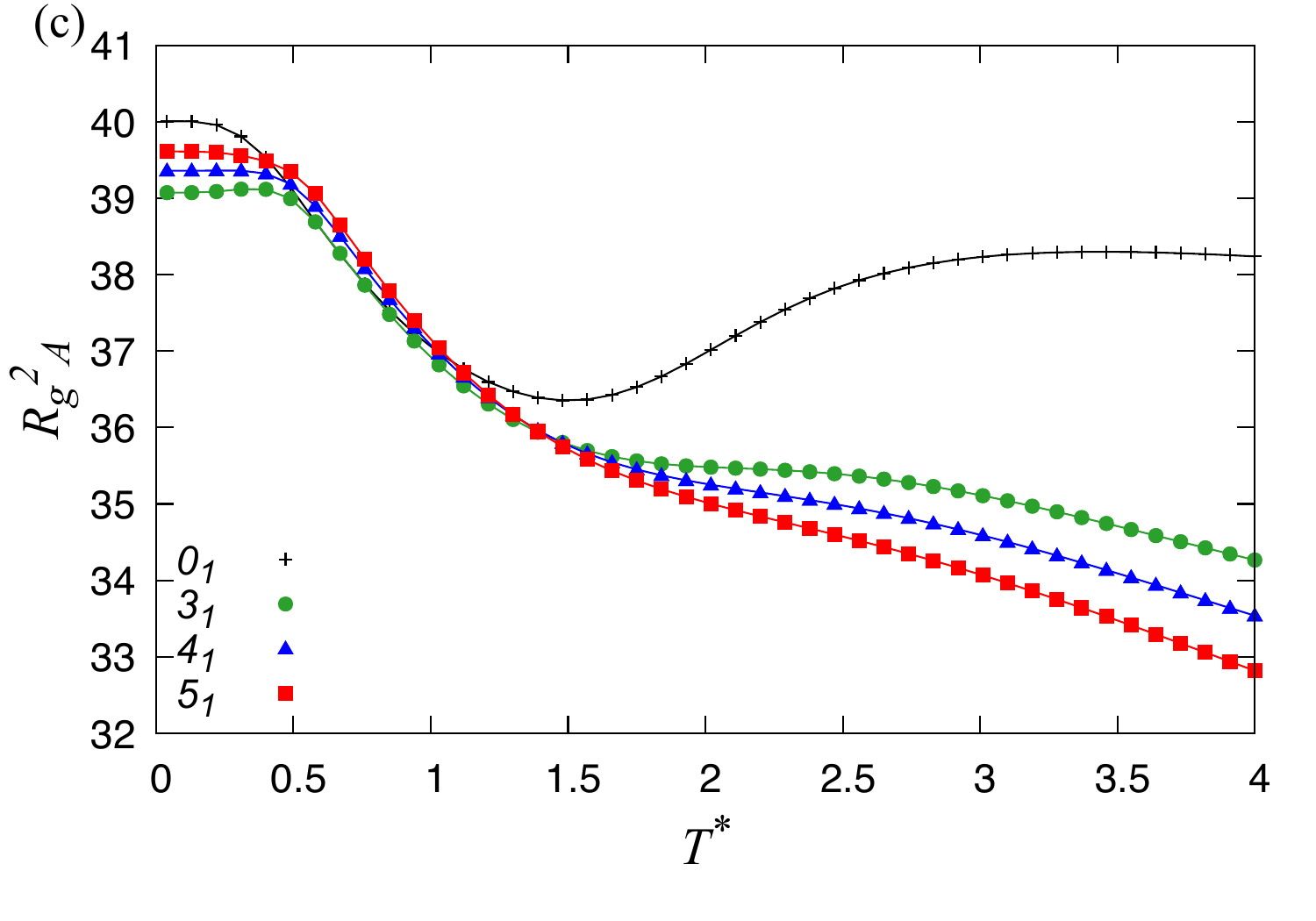}
    \includegraphics[width=0.48\textwidth]{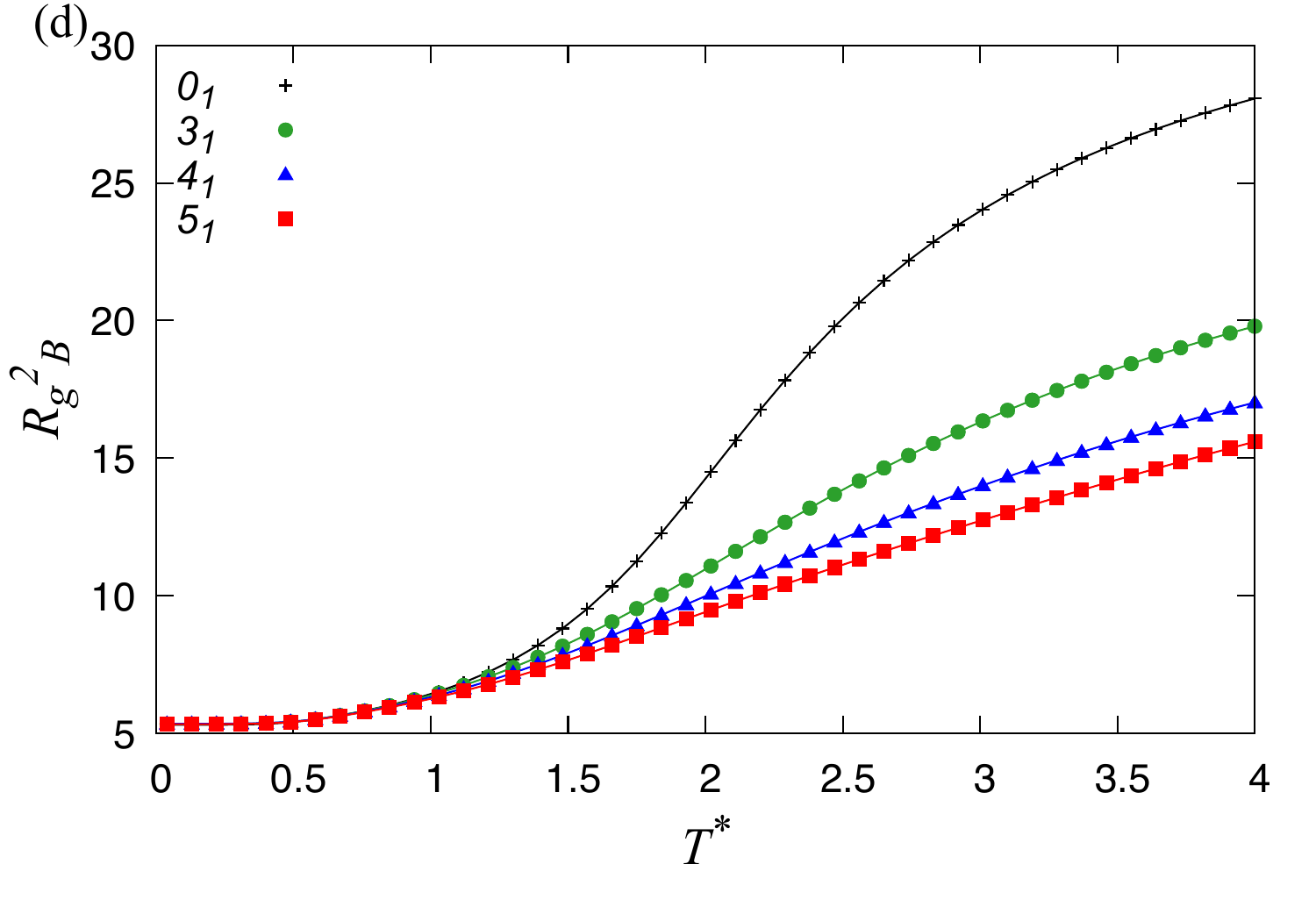} 
    \caption{ \justifying Thermodynamic and structural properties of symmetric diblock
copolymer rings with a symmetric monomer distribution ($N=200$, $N_A=100, N_B=100$) in the AB model for different knot types ($0_1$, $3_1$, $4_1$, and $5_1$). (a) $C/N$ vs $T^*$.
(b) $R_g^2$ vs $T^*$.
(c) $R_{g,A}^2$ vs $T^*$.
(d) $R_{g,B}^2$ vs $T^*$.
}
\label{Topo-100-100-RNA} 
\end{center}
\end{figure}

Fig.~\ref{Topo-100-100-RNA} illustrates the effect of topology on the specific heat capacity and the mean-square radius of gyration of a copolymer ring with a symmetric monomer distribution $N=200$, $N_A=100, N_B=100$. For the unknotted ring ($0_1$), the specific heat capacity exhibits two distinct peaks, whereas for the knotted topologies ($3_1$, $4_1$, and $5_1$), a single peak with a shoulder is observed, Fig.~\ref{Topo-100-100-RNA}(a). This shoulder becomes progressively smoother with increasing topological complexity.

At low temperatures ($T^* \lesssim 0.5$), the radius of gyration of the A block remains nearly constant for all topologies, reflecting a frozen conformational regime. The first peak of the specific heat capacity, occurring at approximately $T^* \simeq 0.6$ for all topologies, coincides with a reduction in $R_{g, A}^2$ (Fig.~\ref{Topo-100-100-RNA}(c)) and partial melting of the B block (Fig.~\ref{Topo-100-100-RNA}(d)). Consequently, the restrictions imposed on the A block are progressively relaxed, leading to an increased number of A--A contacts. 
Although these contacts increase the internal energy, the entropy gain outweighs this energetic cost, leading to a reduction in the overall free energy.
 In the unknotted case, thermal fluctuations at higher temperatures promote an entropic expansion of the A block. Conversely, for knotted topologies, the continued reduction of the A-block size is consistent with knot relocalization effects, which restrict the extension of conformational space through topological constraints. 
Overall, the opposing structural responses of the two blocks largely compensate each other in the case of symmetric monomer distributions with non-trivial topology, leading to a weak dependence of the polymer size on knot type. In contrast, the unknotted copolymer exhibits a stronger expansion with increasing temperature (Fig.~\ref{Topo-100-100-RNA}(b)).	
\subsection{Comparison of AB and HP Models} \label{com}
\begin{figure}
 \begin{center}
    \includegraphics[width=0.48\textwidth]{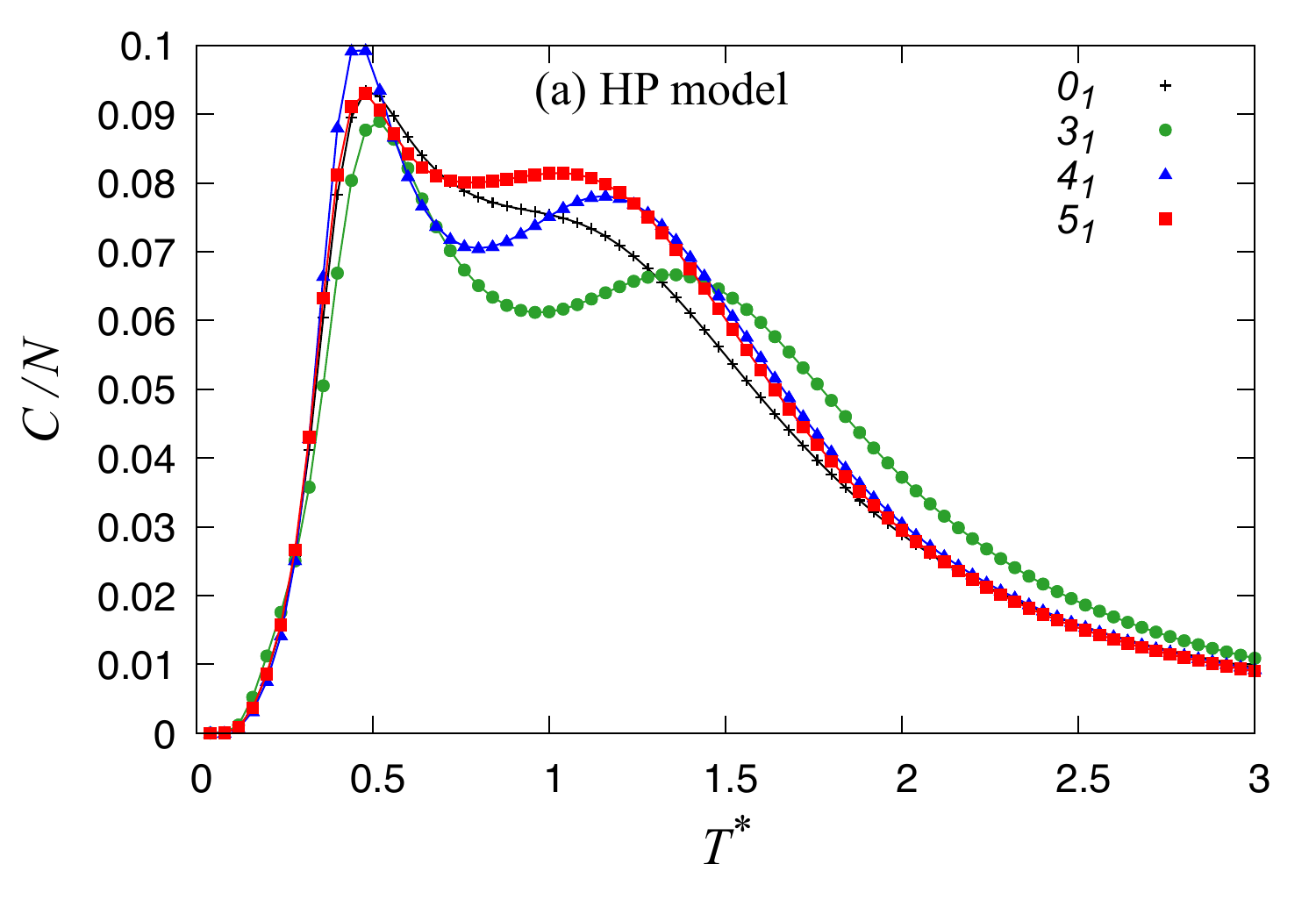}
     \includegraphics[width=0.48\textwidth]{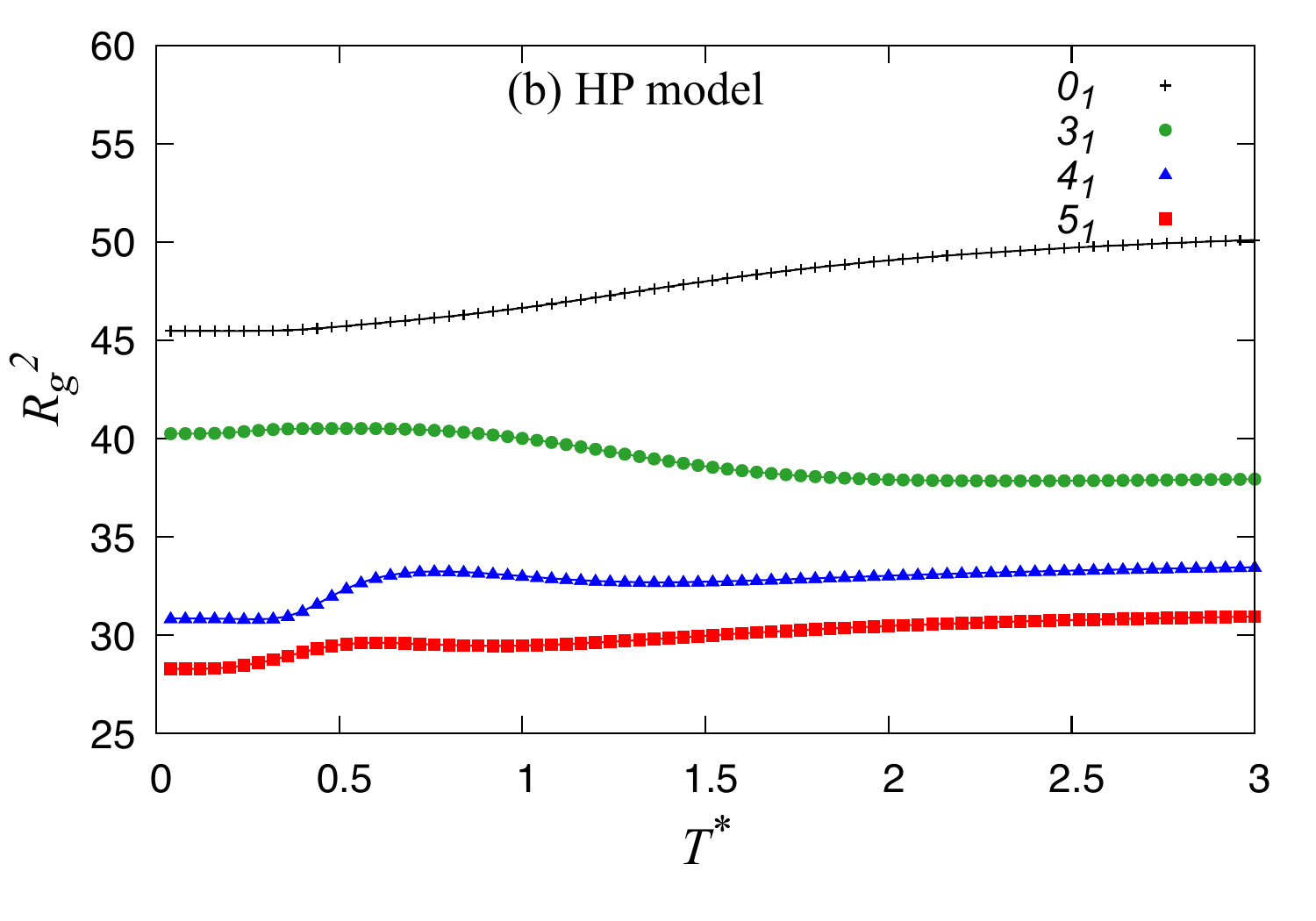}
      \includegraphics[width=0.48\textwidth]{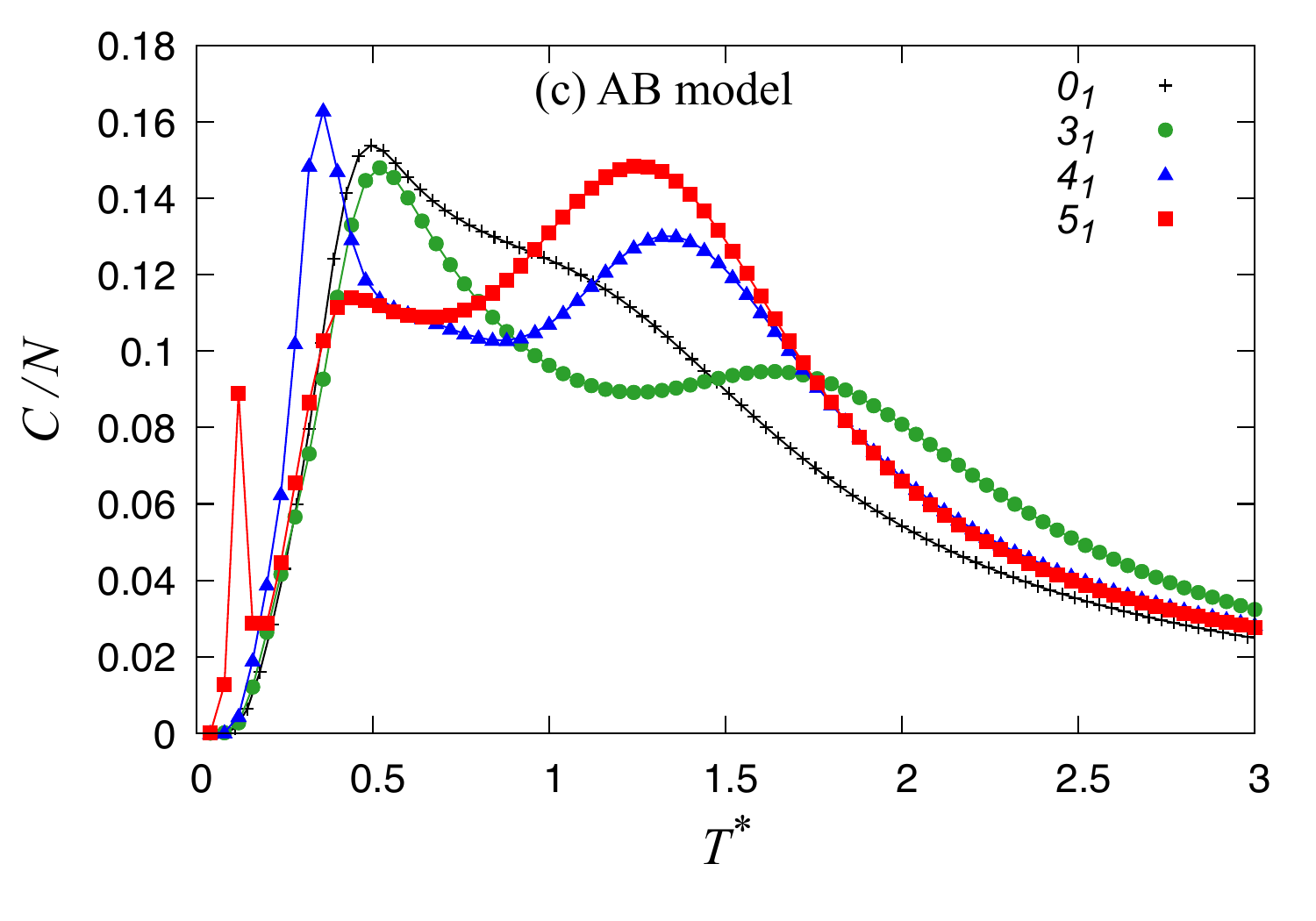}
     \includegraphics[width=0.48\textwidth]{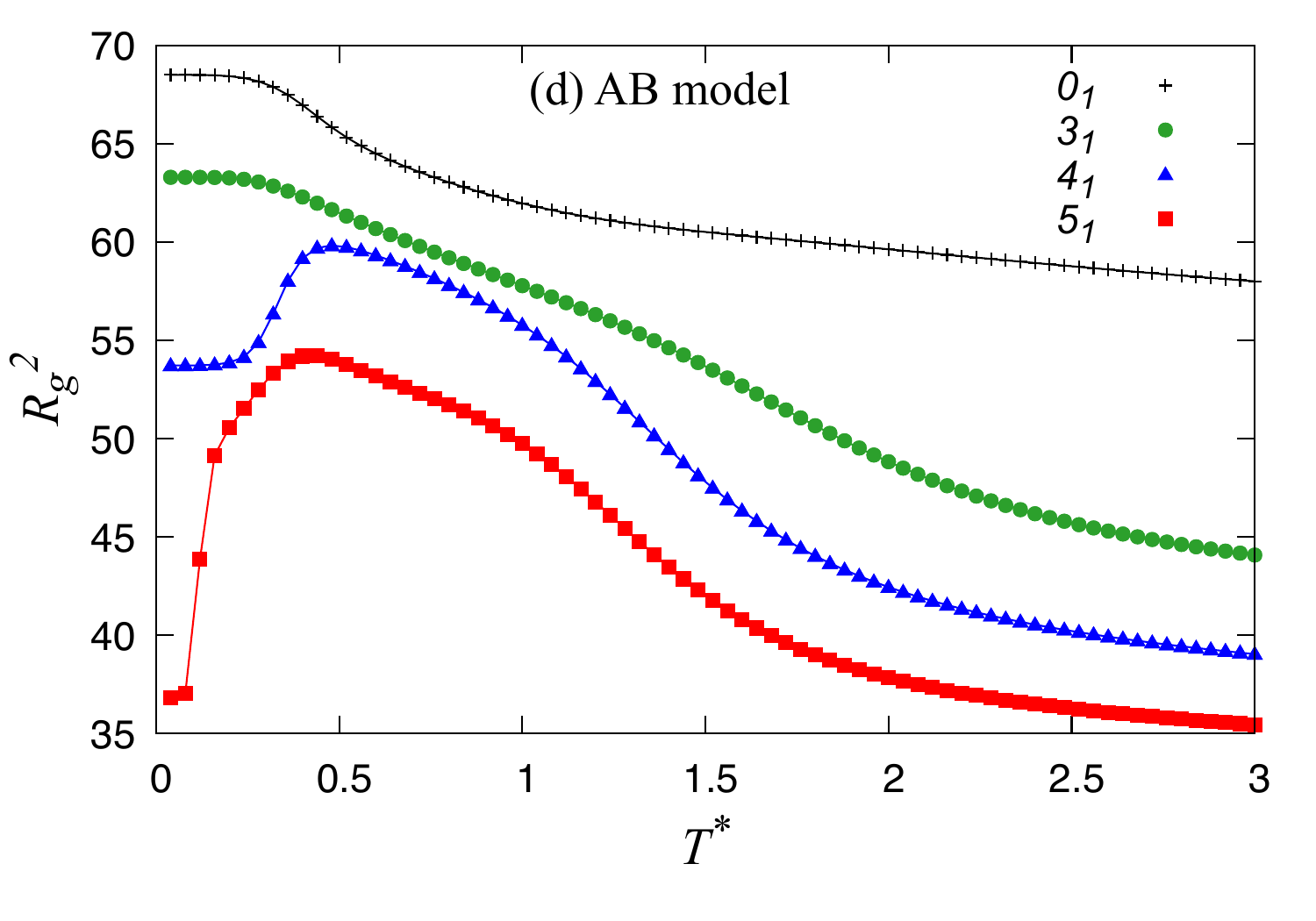}
  \label{Topo-167-33-NNA} 
\caption{\justifying Thermodynamic and structural properties of asymmetric diblock copolymer rings with monomer distribution $(167,33)$ in the HP and AB models for different knot topologies: the unknot $0_1$ and the knots $3_1$, $4_1$, and $5_1$. Panels (a,b) show the specific heat capacity per monomer, $C_V/N$, and the total squared radius of gyration, $R_g^2$, as functions of the reduced temperature $T^*$ for the HP model, while panels (c,d) show the same quantities for the AB model.}
\label{Topo-AB-HP}
\end{center}
\end{figure}

To clarify the role of repulsive monomer--monomer interactions, we compare
asymmetric diblock copolymer rings described by the AB model with their
counterparts in the HP model for several knot topologies, as shown in
Fig.~\ref{Topo-AB-HP}. The comparison is performed for the same monomer distribution
$(167,33)$.

In the HP model (Fig.~\ref{Topo-AB-HP}(a) and ~\ref{Topo-AB-HP}(b)), the radius of gyration exhibits a smooth and nearly monotonic dependence on temperature for all knot types. Although the overall polymer size depends on the knot complexity -- more complex knots yielding smaller radii of gyration -- the qualitative thermal response remains similar across topologies. Minor variations in the position and amplitude of the heat capacity shoulder are observed; however, these do not correlate with pronounced nonmonotonic changes in polymer size.
By contrast, the AB model (Fig.~\ref{Topo-AB-HP}(c) and Fig.~\ref{Topo-AB-HP}(d)) displays a pronounced nonmonotonic temperature dependence of the radius of gyration, most clearly for the $5_1$ knot, whereas a smoother response is observed for the $4_1$ topology. This behavior is characterized by an initial expansion at low to intermediate temperatures, followed by a gradual contraction at higher temperatures.
The pronounced nonmonotonic behavior observed in the AB model, which is absent in the HP case, suggests that the choice of monomer–monomer interactions could enhance topology-dependent size effects. To quantify this effect and identify the regions of monomer composition where topological constraints play a dominant role, we next analyze the maximum size difference between different knot types as a function of temperature and monomer composition.
\subsection{Maximum size difference for a diblock copolymer ring with different topologies}\label {size}
\begin{figure}
  \begin{center}
    \includegraphics[width=0.46\textwidth]{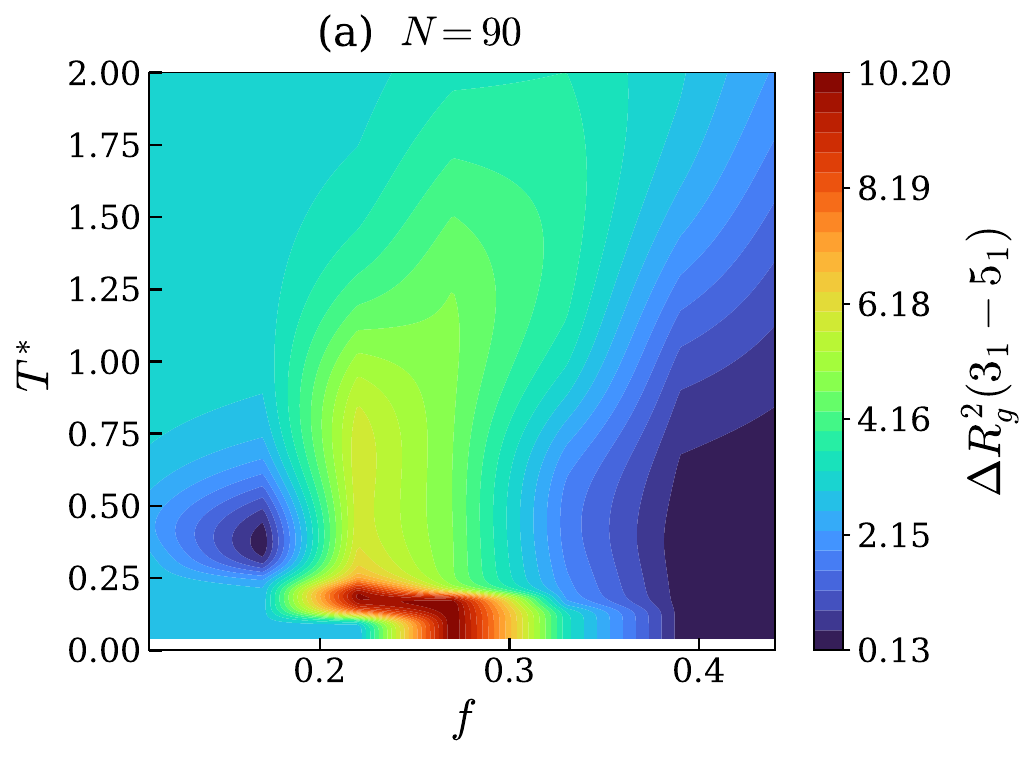}
    \includegraphics[width=0.46\textwidth]{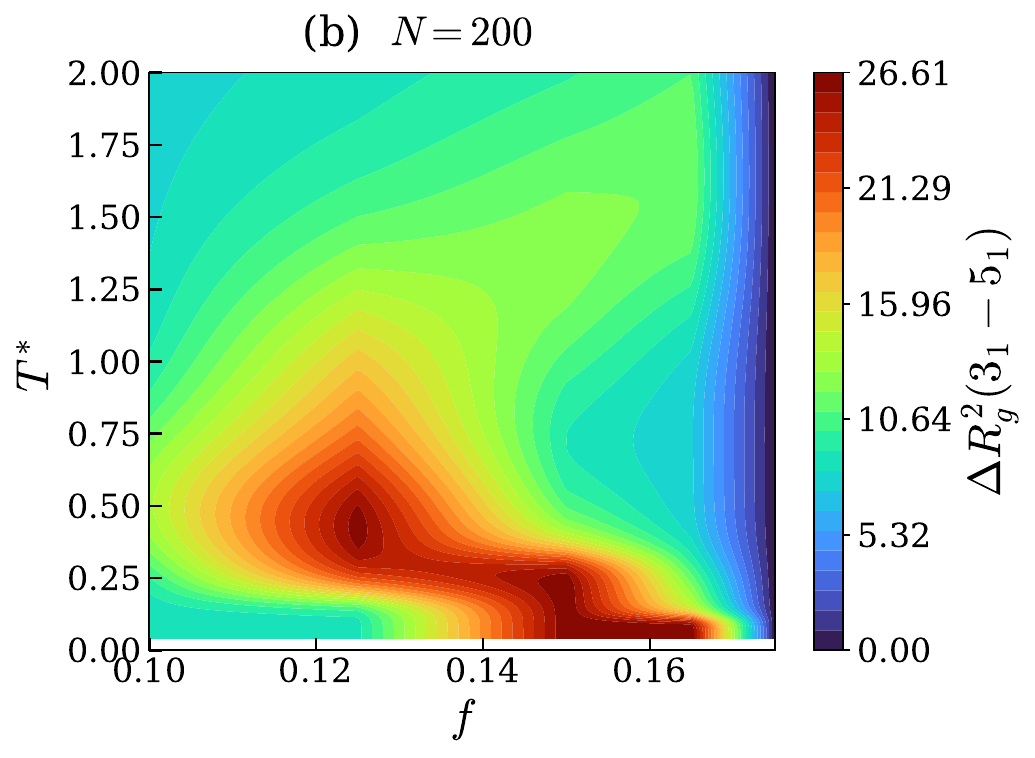}  
\caption{\justifying Color maps of the size difference
$\Delta R_g^2 = R_g^2(3_1) - R_g^2(5_1)$
as a function of temperature $T^*$ and B-block fraction $f$
for diblock copolymer rings of length
(a) $N = 90$ and (b) $N = 200$.
The color scale represents the magnitude of $\Delta R_g^2$,
highlighting regions in the $(f,T^*)$ plane
where the polymer size is most sensitive to knot topology.
For a given chain length $N$, the B-block fraction $f$ corresponds to
$fN$ B-type monomers, allowing direct comparison between different
chain lengths.
}
\label{diff} 
\end{center}
\end{figure}

Color maps of the size difference $\Delta R_g^2 = R_g^2(3_1) - R_g^2(5_1)$ as a function of temperature $T^*$ and B-block fraction $f$ for diblock copolymer rings with total chain lengths $N=90$ and $N=200$ are shown in Fig.~\ref{diff}(a) and (b), respectively. 
The maps are constructed from simulations performed at discrete values of the B-block fraction $f$. 
For $N=90$, $f = 0.11, 0.17, 0.22, 0.27, 0.33, 0.39,$ and $0.44$, 
whereas for $N=200$, $f = 0.10, 0.125, 0.15, 0.165,$ and $0.175$.

The color scale shows how the size difference between the $3_1$ and $5_1$ knots varies with temperature and B-block fraction in the $(T^*,f)$ plane. Regions shown in red indicate large values of $\Delta R_g^2$, where the size difference between the two knot types is most pronounced.

For the longer copolymer ($N=200$), regions of large $\Delta R_g^2$ extend up to temperatures close to $T^* \simeq 1$, whereas for $N=90$ they are limited to temperatures around approximately $T^* \simeq 0.25$. The regions of large $\Delta R_g^2$ are centered at different B-block fractions for the two chain lengths: $f = 0.22$--$0.27$ for $N=90$ and $f = 0.125$--$0.165$ for $N=200$. 
These results demonstrate that the interplay between knot topology, block composition, and temperature strongly influences polymer size differences, with the underlying mechanism arising from the distinct spatial organization of knots rather than from a simple balance of attractive and repulsive interactions.
\section{Conclusions} \label{con}
In this work, we investigate how knot topology influences the thermodynamic and structural properties of finite diblock copolymer rings with heterogeneous interactions. 
Using Wang--Landau Monte Carlo simulations of an AB lattice model with an implicit solvent, we analyze the combined effects of knot type, block composition, and temperature on polymer conformations. The Wang--Landau approach enables efficient sampling at very low temperatures, allowing us to probe the regime in which knot localization induces conformational transitions between compact and extended states.
To characterize these transitions, we compute the heat capacity and radius of gyration over a wide temperature range for copolymer rings with different topologies, including the trefoil ($3_1$), figure-eight ($4_1$), pentafoil ($5_1$), and the unknot.
At lowest temperatures, two distinct regimes are observed depending on knot localization: the knot is either localized within the attractive B block or delocalized over nearly the entire ring. In the first case, with increasing temperature, the knot remains localized within the B block at first, and the polymer undergoes a monotonic shrinkage. When the temperature grows further, the knot delocalizes, and the shrinking continues.
 In the second regime, at the beginning, the knot is spread over the entire ring. Upon heating, the knot localizes within the B block, and  the system exhibits a nonmonotonic, reentrant-like behavior of the radius of gyration, characterized by an expansion followed by a gradual shrinkage upon further heating.
This nonmonotonic behavior is observed for all knot topologies considered ($3_1$, $4_1$, and $5_1$), indicating that it is a general feature of the system. 
However, its occurrence depends on the specific knot type, monomer composition, and chain length, and is encountered in both short and long polymer rings ($N = 90$ and $N = 200$).
The observed size differences among copolymer rings with different knot types at fixed monomer composition reflect topology-dependent differences in knot localization along the chain and arise from the interplay between energetic and entropic effects, with energetic contributions dominating at low temperatures.
Overall, our findings demonstrate that knot topology provides an effective mechanism for tuning the conformational and thermal behavior of diblock copolymer rings in solution. In particular, the degree of block asymmetry ($N_A > N_B$) plays a key role in determining the conformational response of copolymers with different knot topologies.

The present analysis suggests a possible application of knot localization in block copolymers as an energy-storage mechanism. Upon cooling, knot relocation into the B block lowers the system energy and can give rise to low-temperature peaks in the specific heat capacity. The area under these peaks defines a finite energy scale $e_0$, released when the B-block length matches the minimal size required to accommodate the knot.
In systems with multiple B blocks and knots, this mechanism may yield an energy release scaling approximately as $n e_0$, provided each knot localizes within a B block. Although obtained within a lattice model, the existence of a minimal knot size is a general feature of polymer systems. Extending this analysis to off-lattice models is a natural direction for future work.
\begin{acknowledgments} 
The simulations reported in this work were performed in part using the HPC
cluster HAL9000 of
the University of Szczecin.
The research presented here has been supported by the Polish National Science Centre under
grant no. 2020/37/B/ST3/01471.
This work results within the collaboration of the COST
Action CA17139 (EUTOPIA). The use of some of the facilities of the Laboratory of
Polymer Physics of the University of Szczecin, financed by 
a grant of the European Regional Development Fund in the frame of the
project eLBRUS (contract no. WND-RPZP.01.02.02-32-002/10), is
gratefully acknowledged.  \end{acknowledgments}


\begin{thebibliography}{99}
\bibitem{Tubiana2024} L. Tubiana \textit{et al.}, Phys. Rep. \textbf{1075}, 1 (2024).
\bibitem{Fielden} S. D. P. Fielden, D. A. Leigh, and S. L. Woltering, Angew. Chem. Int. Ed. \textbf{56}, 11166 (2017).
\bibitem{Horner} K. E. Horner \textit{et al.}, Chem. Soc. Rev. \textbf{45}, 6432 (2016).
\bibitem{Lim} N. C. H. Lim and S. E. Jackson, J. Phys.: Condens. Matter \textbf{27}, 354101 (2015).
\bibitem{Micheletti} C. Micheletti, D. Marenduzzo, and E. Orlandini, Phys. Rep. \textbf{504}, 1 (2011).
\bibitem{Meluzzi} D. Meluzzi, D. E. Smith, and G. Arya, Annu. Rev. Biophys. \textbf{39}, 349 (2010).
\bibitem{Smerk}R. Staňo, J. Smrek, and C. N. Likos, ACS Nano \textbf{17}, 21369 (2023).
\bibitem{Everaers2004} R. Everaers \textit{et al.}, Science \textbf{303}, 823 (2004).
\bibitem{Kauffman2001} L. H. Kauffman, \textit{Knots and Physics} (World Scientific, 2001).
\bibitem{Chen2016} S. Chen and A. J. Niemi, J. Phys. D: Appl. Phys. \textbf{49}, 315401 (2016).
\bibitem{Janse2011} E. J. Janse van Rensburg and A. Rechnitzer, J. Knot Theory Ramifications \textbf{20}, 1145 (2011).

\bibitem{Lukin} O.~Lukin and F.~V{\"o}gtle, Angew.\ Chem.\ Int.\ Ed.\ \textbf{44}, 1456 (2005).


\bibitem{Ashbridge} Z.~Ashbridge, S.~D.~P.~Fielden, D.~A.~Leigh, L.~Pirvu, F.~Schaufelberger, and L.~Zhang, Chem.\ Soc.\ Rev.\ \textbf{51}, 7779 (2022).
\bibitem{Sauvage} J.-P.~Sauvage and C.~Dietrich-Buchecker, \emph{Molecular Catenanes, Rotaxanes and Knots} (Wiley, 2008).
\bibitem{Fan} X.~Fan, B.~Huang, G.~Wang, and J.~Huang, Macromolecules \textbf{45}, 3779 (2012).
\bibitem{Laurent} B.~A.~Laurent and S.~M.~Grayson, Polym.\ Chem.\ \textbf{3}, 1846 (2012).
\bibitem{Ayme} J.-F.~Ayme, J.~E.~Beves, C.~J.~Campbell, and D.~A.~Leigh, Chem.\ Soc.\ Rev.\ \textbf{42}, 1700 (2013).


\bibitem{Wasserman} S. A. Wasserman and N. R. Cozzarelli, Science \textbf{232}, 951 (1986).
\bibitem{Forte} G. Forte, D. Michieletto, D. Marenduzzo, and E. Orlandini, J. R. Soc. Interface \textbf{18}, 20210138 (2021).
\bibitem{Perlińska} A. P. Perlińska, W. H. Niemyska, B. A. Gren, M. Bukowicki, S. Nowakowski, P. Rubach, and J. I. Sułkowska, Protein Sci. \textbf{32}, e4631 (2023).
\bibitem{Faísca} P.~F.~N.~Fa{\'\i}sca, Comput.\ Struct.\ Biotechnol.\ J.\ \textbf{13}, 459 (2015).
\bibitem{Taylor} W.~R.~Taylor, Nature \textbf{406}, 916 (2000).
\bibitem{Dabrowski} P. Dabrowski-Tumanski and J. I. Sulkowska, Proc. Natl. Acad. Sci. U.S.A. \textbf{114}, 3415 (2017).
\bibitem{marenduzzo} D. Marenduzzo, E. Orlandini, A. Stasiak, D. W. Sumners, L. Tubiana, and C. Micheletti, Proc. Natl. Acad. Sci. U.S.A. \textbf{106}, 22269 (2009).
\bibitem{mallam2006} A. L. Mallam and S. E. Jackson, J. Mol. Biol. \textbf{359}, 1420 (2006).
\bibitem{Jackson2020} S. E. Jackson, Topol. Geom. Biopolym \textbf{746}, 129 (2020).
\bibitem{Majumder2021} S. Majumder \textit{et al.}, Macromolecules \textbf{54}, 5321 (2021).

\bibitem{Tubiana} L. Tubiana, E. Orlandini, and C. Micheletti, Phys. Rev. Lett. \textbf{107}, 188302 (2011).
\bibitem{Orlandini}E. Orlandini and S. G. Whittington, Rev. Mod. Phys. \textbf{79}, 611 (2007).
\bibitem{Zifferer} G.~Zifferer and W.~Preusser, Macromol.\ Theory Simul.\ \textbf{10}, 397 (2001).
\bibitem{Soh} B.~W.~Soh, V.~Narsimhan, A.~R.~Klotz, and P.~S.~Doyle, Soft Matter \textbf{14}, 1689 (2018).
\bibitem{Marcone}B. Marcone, E. Orlandini, A. L. Stella, and F. Zonta, Phys. Rev. E \textbf{75}, 041105 (2007).
\bibitem{Deguchi}T. Deguchi and E. Uehara, Polymers \textbf{9}, 252 (2017).
\bibitem{Caraglio} M. Caraglio, C. Micheletti, and E. Orlandini, Phys. Rev. Lett. \textbf{115}, 188301 (2015).
\bibitem{Kamenetskii} M. D. Frank-Kamenetskii, A. V. Lukashin, and A. V. Vologodskii, Nature \textbf{258}, 398 (1975).
\bibitem{Grzyb2025} A. Grzyb \textit{et al.}, Macromolecules \textbf{58}, 1521 (2025).
\bibitem{Narros}A. Narros, Á. J. Moreno, and C. N. Likos, Macromolecules \textbf{46}, 3654 (2013).
\bibitem{herschberg2021}T.~Herschberg, J.-M.~Y.~Carrillo, B.~G.~Sumpter, E.~Panagiotou, and R.~Kumar,Macromolecules \textbf{54}, 7492 (2021).


\bibitem{Grest} G.~S.~Grest, T.~Ge, S.~J.~Plimpton, M.~Rubinstein, and T.~C.~O'Connor, ACS Polym.\ Au \textbf{3}, 209 (2022).
\bibitem{Wil} R.~J.~Williams, A.~P.~Dove, and R.~K.~O'Reilly, Polym.\ Chem.\ \textbf{6}, 2998 (2015).

\bibitem{Zhao} Y. Zhao and F. Ferrari, J. Stat. Mech. \textbf{2012}, P11022 (2012).
\bibitem{Virnau}P. Virnau, Y. Kantor, and M. Kardar,J. Am. Chem. Soc. \textbf{127}, 15102 (2005).

\bibitem{Wang} W.~Wang, Y.~Li, and Z.~Lu, Sci.\ China Chem.\ \textbf{58}, 1471 (2015).
\bibitem{Kuriata} A.~Kuriata and A.~Sikorski, Macromol.\ Theory Simul.\ \textbf{27}, 1700089 (2018).

\bibitem{Taklim} N.~A.~Taklimi, F.~Ferrari, M.~R.~Pi{\k a}tek, and L.~Tubiana, Phys.\ Rev.\ E \textbf{108}, 034503 (2023).
\bibitem{Taklimi} N.~A.~Taklimi, F.~Ferrari, M.~R.~Pi{\k a}tek, and L.~Tubiana, Phys.\ Rev.\ E \textbf{112}, 045421 (2025).
\bibitem{dill} K.~A.~Dill \emph{et al.}, Protein Sci.\ \textbf{4}, 561 (1995).
\bibitem{Tagliabue21} A.~Tagliabue, C.~Micheletti, and M.~Mella, ACS Macro Lett.\ \textbf{10}, 1365 (2021).
\bibitem{Tagliabue22} A.~Tagliabue, C.~Micheletti, and M.~Mella, Macromolecules \textbf{55}, 10761 (2022).
\bibitem{Ozmaian} M.~Ozmaian and D.~E.~Makarov, PLoS One \textbf{18}, e0287200 (2023).
\bibitem{Enzo} E.~Orlandini, M.~Baiesi, and F.~Zonta, Macromolecules \textbf{49}, 4656 (2016).
\bibitem{copoknotP}L.~Lu, Q.~Qiu, Y.~Lu, L.~An, and L.~Dai, Macromolecules \textbf{57}, 5330 (2024).
\bibitem{Wang2001} F.~Wang and D.~P.~Landau, Phys.\ Rev.\ Lett.\ \textbf{86}, 2050 (2001).
\bibitem{tubiana2018kymoknot} L.~Tubiana, G.~Polles, E.~Orlandini, and C.~Micheletti, Eur.\ Phys.\ J.\ E \textbf{41}, 72 (2018).
\bibitem{Müller} F.~M{\"u}ller-Plathe, ChemPhysChem \textbf{3}, 754 (2002).
\bibitem{van} E.~J.~J.~van Rensburg and A.~Rechnitzer, J.\ Stat.\ Mech.\ P09008 (2011).
\bibitem{Scharein} R.~Scharein \emph{et al.}, J.\ Phys.\ A \textbf{42}, 475006 (2009).
\bibitem{Madras} N.~Madras and A.~D.~Sokal, J.\ Stat.\ Phys.\ \textbf{50}, 109 (1988).
\bibitem{Lai} P.-Y.~Lai, Phys.\ Rev.\ E \textbf{66}, 021805 (2002).
\bibitem{Madras1} N.~Madras and G.~Slade, \emph{The Self-Avoiding Walk} (Springer, 2013).
\bibitem{Ferrari2017} Y.~Zhao and F.~Ferrari, Physica A \textbf{486}, 44 (2017).
\bibitem{Fredrickson} G.~Fredrickson, \emph{The equilibrium Theory of Inhomogeneous Polymers}(Oxford Univ.\ Press, 2005).
\bibitem{Rubinstein} M.~Rubinstein and R.~H.~Colby, \emph{Polymer Physics} (Oxford University Press, Oxford, 2003).
\bibitem{Zhang} G.~Zhang and C.~Wu, Phys.\ Rev.\ Lett.\ \textbf{86}, 822 (2001).
\bibitem{Okay} O.~Okay, Gels \textbf{7}, 98 (2021).
\bibitem{Yong} H.~Yong, Biomacromolecules \textbf{25}, 7361 (2024).
\bibitem{taylor2009phase} M.~P.~Taylor, W.~Paul, and K.~Binder, \textit{J.\ Chem.\ Phys.} \textbf{131}, 114907 (2009).


\end{thebibliography}
\end{document}